\documentclass{aa} 
\usepackage{float} 
\usepackage{graphicx}  
\usepackage{graphics}
\usepackage{txfonts}  
\usepackage{natbib}
\usepackage{adjustbox}
\usepackage{booktabs}
\bibpunct{(}{)}{;}{a}{}{,}
\usepackage[dvipsnames]{xcolor}
\usepackage[normalem]{ulem}
\usepackage{amstext}

\begin{document}  


\title {Glitches in solar-like oscillating F-type stars}
\subtitle{Theoretical signature of the base of the convective envelope on the ratios $r_{010}$}

\author{M. Deal\inst{1,2} \and M.-J. Goupil\inst{3} \and M. S. Cunha\inst{1} \and M. J. P. F. G. Monteiro\inst{1,4} \and Y. Lebreton\inst{3,5} \and S. Christophe\inst{3} \and F. Pereira\inst{1} \and R. Samadi\inst{3} \and A. V. Oreshina\inst{6}   \and G. Buldgen\inst{7} }
  
\institute{Instituto de Astrofísica e Ciências do Espaço, Universidade do Porto,
CAUP, Rua das Estrelas, PT4150-762 Porto, Portugal     
           \and
LUPM, Universit\'e de Montpellier, CNRS, Place Eug\`ene Bataillon, 34095 Montpellier, France
           \and
LESIA, Observatoire de Paris, Universit\'e PSL, CNRS, Sorbonne Universit\'e, Universit\'e de Paris, 5 place Jules Janssen, 92195 Meudon, France
           \and
Departamento de Física e Astronomia, Faculdade de Ciências, Universidade do Porto, Rua do Campo Alegre, 4169-007 Porto, Portugal 
           \and
Université de Rennes, CNRS, IPR (Institut de Physique de Rennes) - UMR 6251, F-35000 Rennes, France
           \and
Sternberg Astronomical Institute, Lomonosov Moscow State University, 119234 Moscow, Russia
           \and
Département d'Astronomie, Université de Genève, Chemin Pegasi 51, CH-1290 Versoix, Switzerland
\\
            \email{morgan.deal@umontpellier.fr} 
            }
           
\date{\today}

\abstract
{The transition between convective and radiative stellar regions is still not fully understood. This currently leads to a poor modelling of the transport of energy and chemical elements in the vicinity of these regions. The sharp variations in sound speed located in these transition regions give rise to a signature in specific seismic indicators, opening the possibility to constrain the physics of convection to radiation transition. Among those seismic indicators, the ratios of the small to large frequency separation for $l=0$ and $1$ modes ($r_{010}$) were shown to be particularly efficient to probe these transition regions. Interestingly, in the Kepler Legacy F-type stars, the oscillatory signatures left in the $r_{010}$ ratios by the sharp sound-speed variation have unexpected large amplitudes that still need to be explained.}
{We analyse the $r_{010}$ ratios of stellar models of solar-like oscillating F-type stars in order to investigate the origin of the observed large amplitude  signatures  of the $r_{010}$ ratios.}
{We tested different possibilities that may be at the origin of the large amplitude signatures using internal structures of stellar models. We then derived an analytical expression of the signature, in particular, of the amplitude of variation, that we tested against stellar models.}
{We show that the signature of the bottom of the convective envelope is amplified in the ratios $r_{010}$ by the frequency dependence of the amplitude compared to the signal seen in the frequencies themselves or the second differences. We also find that with precise enough data, a smoother transition between the adiabatic and radiative temperature gradients could be distinguished from a fully adiabatic region. Furthermore, we find that among the different options of physical input investigated here, large amplitude signatures can only be obtained when convective penetration of the surface convective zone into the underlying radiative region is taken into account. In this case  and even for amplitudes as large as those observed in F-type stars,  the oscillating signature in the r01 ratios can only be detected when the convective envelope is deep enough (i.e. at the end of the main sequence). Assuming that the origin of the large amplitude glitch signal is due to penetrative convection (PC), we find that the PC must extend downward the convective to radiative transition significantly (about $1-2~H_p$) in order to reproduce the large amplitudes observed for the ratios of F-type stars. This deep extension of the convective envelope causes doubt that the origin of the large amplitudes is due to PC as it is modelled here or implies that current stellar modelling (without PC) leads to an underestimation of the size of convective envelopes. In any case, studying the glitch signatures of a large number of oscillating F-type stars opens the possibility to constrain the physics of the stellar interior in these regions.}
{}

  
\titlerunning{}
  
\authorrunning{}  

\maketitle 


\section{Introduction}

\begin{figure*}
    \centering
    \includegraphics[scale=0.6]{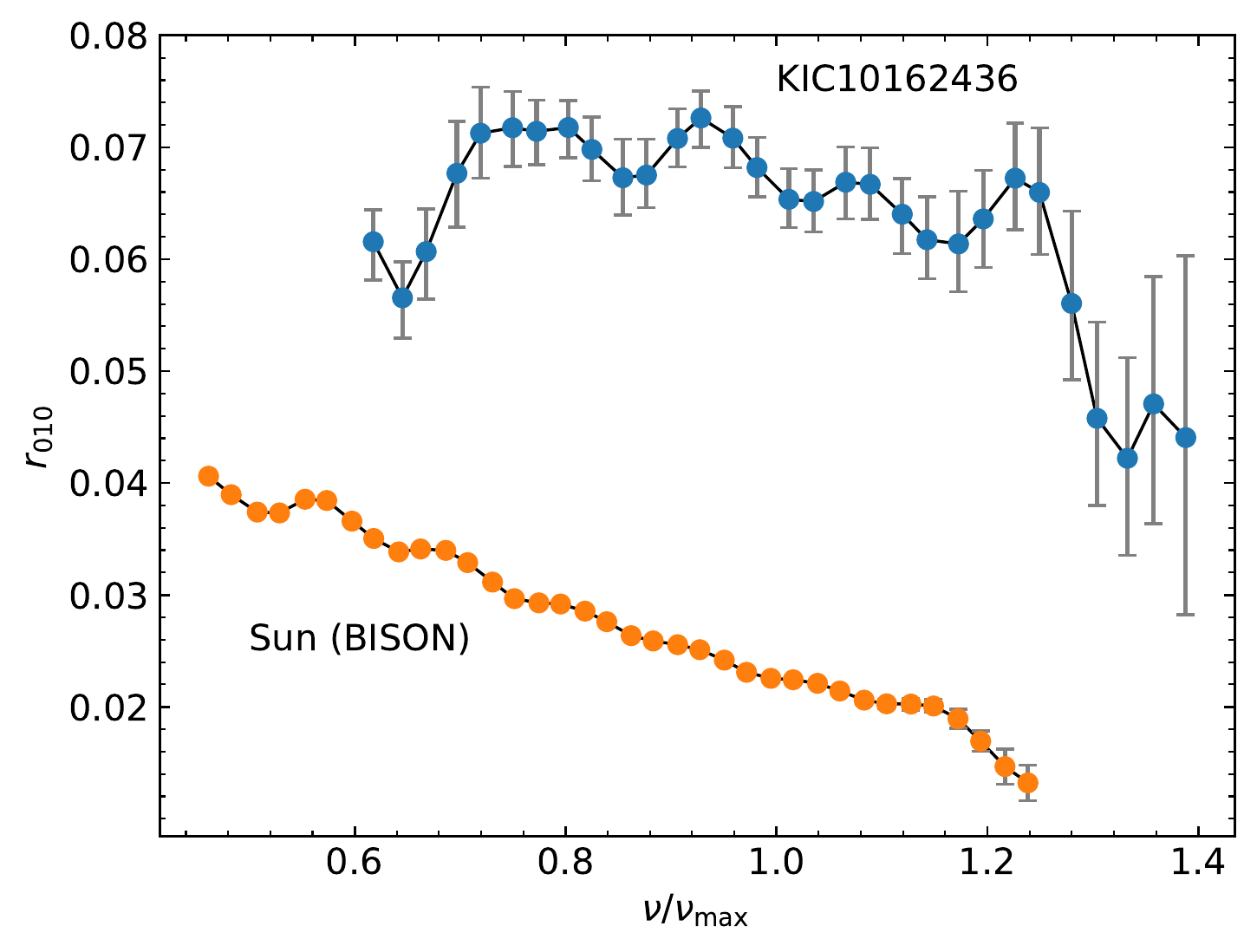}
    \includegraphics[scale=0.6]{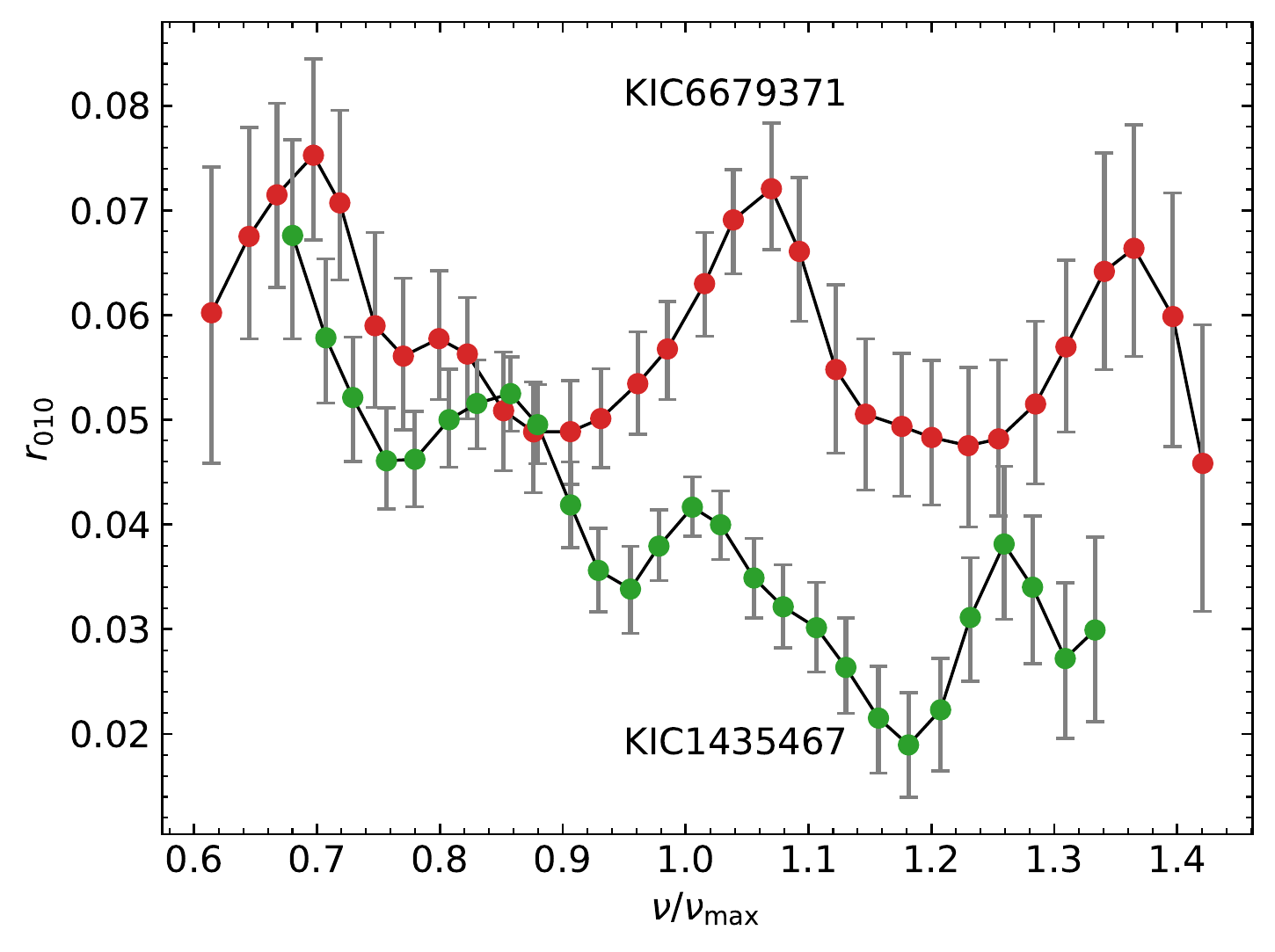}
   \caption{$r_{010}$ ratios according to the frequency scaled to $\nu_\mathrm{max}$. \textbf{Left}: For the Sun (computed from the BISON frequencies; \cite{broomhall09,hale16}) and KIC10162436. \textbf{Right}: For KIC1435467 and KIC6679371 (data from \citealt{lund17}).}
   \label{fig:1}
\end{figure*}

Asteroseismology is a powerful tool for probing stellar interiors. This is done through the interpretation of stellar intrinsic oscillation frequencies \citep[e.g.][and references therein]{cox80,unno89,gough93,cunha07,chaplin13,garcia19}. The properties of the oscillation modes and their frequencies are determined by the structure in the regions in which they are trapped. In solar-like oscillating stars such as the Sun, the turbulent convection in the outer layers excites the normal modes of the star in a frequency domain in which they are almost purely acoustic, and their frequencies follow a nearly regular pattern (asymptotic regime). 
Deviations from this regularity arise when the structure varies more rapidly locally than the wavelength of the excited modes (so-called glitch). This generates an oscillatory component (so-called glitch signature) to the frequencies and the frequency differences in addition to the smooth component \citep{gough90,roxburgh01}. The study of the signatures of acoustic glitches has been proven to successfully unveil internal stellar characteristics such as the location of the base of the surface convective zone (BSCZ) and the helium second-ionisation (HeII) zone for the Sun \citep{monteiro94,roxburgh94,JCD95,ballot04,monteiro05,houdek07,roxburgh09,JCD11} and distant stars \citep{monteiro00,gough02,mazumdar12,verma17,verma19,farnir19}.

Here we focus on the determination of the location of the base of the convection zone and its properties. The determination of the location of the BSCZ  allows us to characterise the convection process at the transition with the underlying radiative interior. This is crucial to improve our current understanding of stellar convection, and especially to constrain its modelling in stellar evolution models. The current modelling of convection in stellar evolution models is based on the mixing length theory (hereafter MLT) \citep{bohm58,canuto96}. This theory assumes that the fluid acceleration vanishes at the boundary of the convective zone, defined either by the Schwarzschild or Ledoux criteria. However, due to their inertia, convective cells penetrate the sub-adiabatic region below the convective envelope. These motions are able to impact the local temperature gradient and mix the chemical elements to an extent that is not yet known. Three types of processes can occur in this situation \citep[e.g.][]{zahn91,anders22}. Convective overshoot corresponds to turbulent motions from the convective region which penetrate in the neighbouring radiative region but not altering the radiative gradient. Convective entrainment refers to turbulent motions which do alter the radiative gradient or erode the gradient of chemical composition. The third process, the penetrative convection, changes the radiative temperature to the adiabatic one in a region that is stable by the Schwarzschild criterion \citep[see][for a more detailed description of the three processes]{anders22b}. Penetrative convection is the focus of this work. \cite{zahn91} predicted a penetrative convection region (i.e. with a fully adiabatic temperature gradient) at the base of the convective envelope of the order of one pressure scale height. The extension of the penetrative region was measured for the Sun to be about 0.2-0.3 pressure scale height ($H_p$). In addition, a smooth and mostly subadiabatic transition from the adiabatic to a radiative temperature gradient was favoured by the data \citep[e.g.][]{JCD11,zhang19}.  

Different seismic indicators can provide information on the location of the BSCZ: the frequency variations can be inspected directly \citep[e.g.][]{monteiro94}, or alternatively, quantities such as the phase shift variation \citep{roxburgh94,roxburgh09}, the second differences \citep[e.g.][]{gough90}, the fourth differences \citep{basu97}, and the ratios of small to large separation \citep{roxburgh09,mazumdar14}. Because the amplitude of the signature of the acoustic glitches is a few orders of magnitude lower than the actual frequencies, all these methods require high-precision data. This precision was achieved for distant stars with the space missions CoRoT \citep{baglin06,baglin13} with uncertainties of about $0.50~\mu$Hz, and with \textit{Kepler} \citep{borucki09,koch10} with an uncertainty of about $0.05~\mu$Hz for the best targets.
With the exception of the ratios of the small to large separations, in all seismic indicators mentioned above, the signature of the helium second-ionisation region dominates the signature left by the BSCZ. This issue can be critical when the uncertainties on the individual frequencies are large, especially for F-type stars. \cite{monteiro94} developed a method, using the frequency variation, that allows filtering the HeII signature out. This method was successful in determining the location of the BSCZ of the Sun and other \textit{Kepler} stars \citep{pereira17}. Later, \cite{roxburgh09} showed that the signature of the BSCZ dominates in the ratios of small to large separations of $l=0,1$ p-modes (hereafter $r_{010}$) of the Sun, while the signature of HeII is found to be residual. The frequency ratios can be described by a smooth variation in frequency (hereafter a smooth component), on which is superimposed a periodic variation with frequency (oscillating component). The smooth component is an efficient indicator of the convective core extension \citep{deheuvels16}.

Similarly to the Sun, the oscillatory component found in the $r_{010}$ ratios of the CoRoT star HD52265 was proposed to be the signature of envelope penetrative convection, hereafter PC \citep{lebreton12,lebreton14}. This was the first detection of this signature in another star than the Sun. Moreover, the amplitude of the signature was found to be larger than in the Sun, even though the stars have similar properties. \cite{mazumdar14} used the oscillatory signature in the $r_{010}$ ratios to determine the acoustic radius of the BSCZ of 12 \textit{Kepler} stars. The use of the $r_{010}$ ratios to probe PC was extended to 19 \textit{Kepler} Legacy stars by \cite{christophe19}. The amplitude of the signature for some stars was found to be much larger than the one found for the Sun, allowing us to characterise the location of the BSCZ for distant stars, despite the larger uncertainties in the frequencies. 
 
In the present work, we focus on the ratios of small to large separations $r_{01}$ and $r_{10}$ ($r_{010}$), and more precisely, on the oscillating part of these ratios. Our goal is to investigate whether the origin of the large-amplitude signature in the oscillations of these ratios in F-type stars is PC. Then, we investigate to what extent constraints may be placed on the temperature gradient immediately below the convective envelope and on the extension of the PC region if this process caused the signature. Our study is based on stellar models and on an analytical (nearly independent of the stellar model) study of this signature in the sharply varying region (adapted from \citealt{monteiro94}). 

In Section 2 we describe the seismic indicators used to characterise the location of the BSCZ and the observations. In Section 3 we investigate the origin of the large amplitude signature in solar-like oscillating F-type stars. In Section 4 we define an expression of the signature of the glitch in the ratios according to the structure of the star in the sharply varying region. Then, we compare theoretical predictions of the amplitude and period of the signal to those of stellar models in Section 5. We determine the detectability of this signature in Section 6. In Section 7 we compare the results obtained with the frequency ratios to those obtained with other seismic indicators, namely the frequencies themselves and the second differences. The impact of some specific changes in the structure below the BSCZ is addressed in Section 8. We finally discuss the results and conclude in Section 9.

\section{Large-amplitude signature in the ratios $r_{010}$ of solar-like oscillating F-type main-sequence stars}

\subsection{Definition}

The five-point small separations between modes (with frequencies $\nu_{n,l}$) of angular degree $l=0$ and $1$ and radial order $n$ are defined by \citep{roxburgh03}
\begin{equation}
    d_{01}(n)=\frac{1}{8}(\nu_{n-1,0}-4\nu_{n-1,1}+6\nu_{n,0}-4\nu_{n,1}+\nu_{n+1,0}),
\end{equation}
\begin{equation}
    d_{10}(n)=-\frac{1}{8}(\nu_{n-1,1}-4\nu_{n,0}+6\nu_{n,1}-4\nu_{n+1,0}+\nu_{n+1,1}).
\end{equation}

The ratios of small to large separation are defined by 
\begin{equation}\label{eq:r010}
    r_{01}(n)=\frac{d_{01}(n)}{\Delta_1(n)}~~\mathrm{and}~~r_{10}(n)=\frac{d_{10}(n)}{\Delta_0(n+1)},
\end{equation}
\noindent where $\Delta_l(n)=\nu_{n,l}-\nu_{n-1,l}$ is the large separation. We hereafter refer to $r_{010}$ as the function representative of $r_{01}$ and $r_{10}$ with frequency (see the curves of Fig.~\ref{fig:1}). These ratios of small to large separations are only weakly sensitive to surface layers \citep{roxburgh03}, which means that surface effects do not need to be considered when observations are compared to models. Magnetic activity may nevertheless have a slight impact on the ratios \citep{thomas21}. For stars similar to the Sun, these specific ratios are dominated by the signature of the BSCZ and convective core, and they are less sensitive to the helium second-ionisation glitches. On the other hand, these frequency differences have larger uncertainties than the frequencies alone, and care must be taken when the observed ratios are interpreted. 

\begin{table}
\centering
\caption{Inputs physics of the A0 model.} 
\label{tab:A0}
\begin{tabular}{l|c}
\noalign{\smallskip}\hline\hline\noalign{\smallskip}
Mass (M$_{\odot}$) & 1.40 \\
Opacities & OP \\
X$_{ini}$ & 0.7231  \\ 
Y$_{ini}$ & 0.2620 \\
(Z/X)$_{ini}$ & 0.0206 \\
$\alpha_\mathrm{CGM}$ & 0.6838  \\ 
Atmosphere & Eddington \\
Metal mixture & AGSS09 + S10  \\ 
Equation of State & OPAL2005  \\
Nuclear reaction rates & NACRE + LUNA \\ 
Core overshoot ($H_p$) & 0.15 \\
Transport & Atomic diff. (without rad. acc.) \\ 
\noalign{\smallskip}\hline\noalign{\smallskip}
\end{tabular}
\end{table}
\begin{table*}[!ht]
\centering
\caption{ Stellar models with  changes in the input physics with respect to model A0. } 
\label{tab:1}
\begin{tabular}{c|l|l}
\hline\hline
Model & Affected & Change in the input physics\\
\hline
A0 & - & - \\
~A1 & $\Gamma_1$ & \textbf{EoS}: SAHA-S\tablefoottext{a}, \textbf{Turbulent mixing}: calibrated on helium surface abundance of \textit{Kepler} F-type stars \tablefoottext{b}\\
A2 & $\mu$ &\textbf{Turbulent mixing}: homogenising turbulent mixing down to T=$5\times10^6$~K, and no mixing below\\
A3 & $T$ & \textbf{Surface penetrative convection}: $\xi_{PC}$\tablefoottext{c}$=0.2$\\
A4 & $T$ & \textbf{Surface penetrative convection}: $\xi_{PC}$\tablefoottext{c}$=1.0$\\
A5 & $T$ & \textbf{Surface penetrative convection}: $\xi_{PC}$\tablefoottext{c}$=2.0$\\
A6 & $T$ & \textbf{Surface penetrative convection}: Extension equivalent to model A5 with a smoother transition ($\beta=0.5$)\\
A7 & $T$ & \textbf{Surface penetrative convection}: Extension equivalent to model A5 with a smoother transition ($\beta=2.0$)\\
\hline
B1 & $T$ & \textbf{No atomic diffusion}\\
B2 & $T$ & \textbf{No atomic diffusion}, \textbf{Penetrative convection}: $\xi_{PC}$\tablefoottext{c}$=2.0$\\
B3 & $T$ & \textbf{No atomic diffusion}, \textbf{PC}: Extension equivalent to model B2 with a smoother transition ($\beta=0.5$)\\
B4 & $T$ & \textbf{No atomic diffusion}, \textbf{PC}: Extension equivalent to model B2 with a smoother transition ($\beta=2.0$)\\
\hline
\end{tabular}
\tablefoot{\tablefoottext{a}{\cite{baturin17}}. \tablefoottext{b}{\cite{verma19b}, we used a turbulent diffusion coefficient because of the strong helium and metal depletions induced by atomic diffusion in F-type stars. This enables us to stay in the parameter space of the equation of state (EoS) for the hydrogen mass fraction ($0.10<X<0.90$)}. \tablefoottext{c}{\cite{zahn91}}. }
\end{table*} 

\subsection{Observations}

Figure~\ref{fig:1} shows the $r_{010}$ ratios according to the ratio of the frequency $\nu$ and the frequency at maximum power $\nu_\mathrm{max}$ for the Sun, computed from the BISON frequencies \citep{broomhall09,hale16} and for three \textit{Kepler} Legacy stars (KIC10162436, KIC6679371 and KIC1435467) using the ratios provided by \cite{lund17,lund17er}. Two main characteristics can be seen: a long-term smooth trend, and an oscillatory component. The long-term trend of the ratios decreases with increasing frequencies. It has been shown that the slope and mean value of $r_{010}$ ratios are good indicators of the extension of convective cores (when present) and of the amount of hydrogen in the core \citep{popielski05,cunha07a,deheuvels10,aguirre11,cunha11,brandao14,deheuvels16}.

The oscillatory component is visible in the ratios of all stars presented in Fig.\ref{fig:1}. For the Sun, \cite{roxburgh09} showed that this signature is related to the BSCZ and that the period of the signal is related to its acoustic radius $t(r)=\int_0^rdr/c$ (its counterpart, the acoustic depth $\tau$, is defined by $\tau(r)=\int_r^{R^\ast}dr/c$, with $R^\ast$ being the radius of the star). \cite{christophe19}\footnote{See chapter II of https://tel.archives-ouvertes.fr/tel-02883979/document} showed that these oscillations are also present in some \textit{Kepler} Legacy stars (at least 19 stars with amplitudes larger than the uncertainties on the ratios). He identified the signal to be the signature of a glitch below the second helium-ionisation zone and associated it with the BSCZ. He also showed that for stars in which the signature can be measured, the amplitude of the signal is larger for F-type than for G-type stars. In the following sections, we assess the origin of these large signals compared to the Sun and determine the link between the shape of the signal and the physical properties of the sharply varying region.

\section{Origin of the large-amplitude signals}

\subsection{Possible causes}

The oscillatory component of the ratios $r_{010}$ or the signature of the acoustic glitch is induced by a sharp variation in the adiabatic sound speed,
\begin{equation}
    c_s^2=\frac{\Gamma_1P}{\rho}\approx\frac{k_B\Gamma_1 T}{m_H\mu},
\end{equation}
\noindent where $\Gamma_1$ is the first adiabatic exponent, $P$ is the pressure, $\rho$ is the density, $T$ is the temperature, $\mu$ is the mean molecular weight at a given level within the star, $k_B$ is the Boltzmann constant, and $m_H$ is the proton mass. The rapid variation in the sound speed can arise from sharp variations in $\Gamma_1$, the temperature, and/or in the mean molecular weight $\mu$, or from a variation of all these parameters.

However, as we show in Sect.~\ref{s:input}, standard stellar models do not predict an oscillatory component with large amplitudes. Hence the origin of the large magnitude of the oscillation amplitude is not yet identified. Three cases are then possible:

\textit{First adiabatic exponent} ($\Gamma_1$): A sharp variation in $\Gamma_1$ may be induced by an ionisation region that is not well taken into account in stellar models, such as the region of heavy elements, which is located below the helium region. \cite{brito17,brito18,brito19} showed that these ionisation regions create a signal in the derivative of the frequency phase shift. This signature is not well reproduced by the models, and the authors suggested that a process such as radiative accelerations \citep[e.g.][]{michaud15} could maintain enough metals in the surface convective zone to reconcile models and observations. These authors also suggested that an equation of state based on the chemical approach, such as SAHA-S \citep{gryaznov04, ayukov04, gryaznov06, baturin13,baturin17}, may induce this type of signature in stellar models. 

\textit{Mean molecular weight} ($\mu$): Sharp variations in $\mu$ occur at the transition between fully mixed and stable regions (i.e. at the transition between convective and radiative regions) and are usually neglected in standard models. 

\textit{Temperature} ($T$): Sharp variations in the temperature gradient are due to the transition between different processes of energy transport inside the star due to the presence of convection. They are known to induce glitches, especially in presence of penetrative convection or overshooting. Because the amplitude of the oscillatory component in the $r_{010}$ ratios is larger for F-type stars, the signal may also come from the iron-nickel convective zone induced by the accumulation of these elements by the effect of radiative accelerations around $T\approx200~000$~K \citep{richard01,theado09,deal16}. This iron/nickel convective zone appears deeper than the hydrogen/helium zones.

Frequency ratios $r_{010}$ have little sensitivity to the surface layers, and some of the above possibilities can therefore be discarded. The impact of radiative accelerations is mostly noted at the very surface of stars, and we tested that the impact of the accumulation of heavy elements and the formation of an iron convective zone cannot be detected by the ratios (see Sect.~\ref{detect} for more details). Competition between microscopic and macroscopic transport processes is also likely to reduce the formation of a strong mean molecular weight gradient.

\subsection{Stellar models}\label{s:input}

In order to identify the origin of the large-amplitude oscillatory component of the $r_{010}$ ratios, we computed stellar models with the Cesam2k20 stellar evolution code \citep[the new version of the Code d'Evolution Stellaire Adaptatif et Modulaire, previously called CESTAM;][]{morel08,marques13,deal18} with the aim of testing different scenarios according to the possible causes mentioned above. The input physics of the reference standard model A0 is presented in Table~\ref{tab:A0}. The other models are presented in Table~\ref{tab:1}.

\subsubsection{Input physics}

All stellar models were computed with a mass of $1.40$~M$_\odot$, typical of the F-type stars at solar metallicity, and $X_c<0.3,$ which is representative of the three F-type stars shown in Fig.~\ref{fig:1} (see \citealt{silva17}). The effect of mass is addressed in Section~\ref{detect}. We used the OPAL2005 \citep{rogers02} or SAHA-S \citep{baturin17} equations of state and the OP opacity tables \citep{seaton05}. Nuclear reaction rates were taken from the NACRE compilation \citep{angulo99}, except for the $^{14}\mathrm{N}(p,\gamma)^{15}\mathrm{O}$ reaction, for which we used the LUNA rate \citep{imbriani04}. We adopted an Eddington grey atmosphere for surface boundary conditions. Convection was computed according to the \cite{canuto96} formalism with a solar-calibrated mixing length parameter, $\alpha_\mathrm{CGM}=0.6838$. We chose the AGSS09 metal mixture \citep{asplund09} with meteoritic abundances for refractory elements from \cite{serenelli10} and adopted calibrated solar values for the initial composition. All stellar models, except when specified otherwise, include the effect of atomic diffusion without radiative accelerations following the \cite{michaud93} formalism. 

Convective-core step overshoot was included with an extension of $0.15~H_p$ except when specified otherwise. Anticipating the results of later sections, we draw specific attention to the impact on the ratios of the temperature gradient profiles below the convective envelope that is induced by convective penetration.

\subsubsection{Temperature gradient profiles in the PC region}\label{nablaT}

We considered two types of temperature gradients in the penetrative convective region below the convective border defined by the Schwarzschild criterion. Models A3, A4, and A5 include a PC region with a fully adiabatic temperature gradient following the \cite{zahn91} formalism, whereas models A6, A7, B3, and B4 were built assuming a smoother transition between the adiabatic and the radiative temperature gradients, following Eq.~\ref{eq:smooth_nablaT} (qualitatively similar to the temperature gradient shown in Fig.~2 of \citealt{anders22}). In all these models, the PC region is assumed to be fully mixed with a uniform chemical composition.\\

\textit{Fully adiabatic temperature gradient.}
The \cite{zahn91} formalism is based on two main hypotheses. Firstly, the temperature gradient is considered fully adiabatic in the PC region. Secondly, the convective flux is assumed to be proportional to $\rho^3W$, where $W$ is the root mean square of the velocities directed downward. This leads to an extent of the penetrative convective region ($L_p$) defined by
\begin{equation}
    \frac{L_p}{H_p}=\frac{\xi_{PC}}{\chi_p},
\end{equation}
\noindent where $H_p$ is the pressure scale height, $\chi_p$ is the conductivity gradient, and $\xi_{PC}$ is the ratio of the convective efficiency in the unstable region to that of the stable region. This last parameter cannot be determined from first principles and needs to be calibrated. \cite{zahn91} determined that $\xi_{PC}$ should have a value of the order of unity \citep[see also][for the Sun]{berthomieu93}. The higher its value, the deeper the PC region. Models A3, A4, and A5 were computed with $\xi_{PC}=0.2$, $1.0,$ and $2.0$, respectively. This represents $3.9$, $19,$ and $39$\% of the size of the surface convective zones of the models at $X_c=0.10$, respectively. For a $1.0$~M$_\odot$, assuming the same input physics as model A4 ($\xi_{PC}=1.0$), we find an increase of $21$ and $22$\%  of the depth of the convective envelope at the age of the Sun and at $X_c=0.10$, respectively. However, for the Sun, the increase is found to be about ten times smaller from helioseismology \citep[e.g.][]{JCD11,zhang19}.

We stress that $\xi_{PC}\approx1.0$ found for the Sun in the early 1990s came for the use of an equation of state that was less accurate than the current ones. This  affects the value of $\chi_P$, hence the value of $\xi_\mathrm{PC}$. All this tends to indicate that the value of $\xi_\mathrm{PC}$ depends on the input physics of the models and therefore probably on the type of stars. Hence, the same calibration value should not be expected for $\xi_\mathrm{PC}$ for an F-type and a G-type star.\\

\textit{Smoother temperature gradient in the PC region.}
For the Sun, \cite{baturin10} proposed that the temperature gradient in the PC region undergoes a smooth transition between the adiabatic and radiative gradient. Later, \cite{JCD11} confirmed this with helioseismology. \cite{anders22} showed with 3D simulations that the penetrative convective regions were adiabatic over almost 90\% of the region, with a smoother rather than a steep transition close to the bottom of the region. They also predicted a PC region extension of 20-30\% of a mixing length, which is larger than the depth seismically characterised for the Sun.

When the large amplitude of the glitch signatures in F stars is assumed to be due to the presence of a PC region below the CZ, PC extensions larger than that of the Sun must be considered. Accordingly, we cannot use the expression of the temperature gradient from \cite{JCD11}. This expression assumes that the radiative gradient decreases monotonically below the Schwarzschild boundary, which is not the case for the stellar envelopes in F-type stars. Following the shape of the transition predicted by \cite{anders22} (see their Fig.~2), we rather define an empirical ad hoc expression for the temperature gradient in the PC region,
\begin{equation}\label{eq:smooth_nablaT}
   \nabla= \nabla_\mathrm{ad} - \frac{\nabla_\mathrm{ad}- \nabla_\mathrm{rad}}{2}\left[1 - \frac{2}{\pi}\arctan\left(\frac{\zeta(r)-\alpha_{PC}H_p(r_\mathrm{cz})}{\beta(\nabla_\mathrm{ad}-\nabla_\mathrm{rad})^4}\right)\right],
\end{equation}

\noindent where $\zeta(r)=r-r_\mathrm{cz}$, with $r_{cz}$ being the radius where $\nabla_\mathrm{ad}=\nabla_\mathrm{rad}$, $\alpha_{PC}$ is the extension of the PC region in units of pressure scale height, $H_p(r_\mathrm{cz})$ is the pressure scale height at $r_\mathrm{cz}$ , and $\beta$ is the parameter controlling the steepness of the transition (higher values of $\beta$ lead to smoother transitions). 
Models A6 and B3 were computed with $\beta=0.5$ and $\alpha_\mathrm{OV}=1.0~H_p$, and models A7 and B4 were computed with $\beta=2.0$ and $\alpha_\mathrm{OV}=1.0~H_p$.

\subsubsection{Cause of the signal}

\begin{figure}
    \centering
    \includegraphics[scale=0.6]{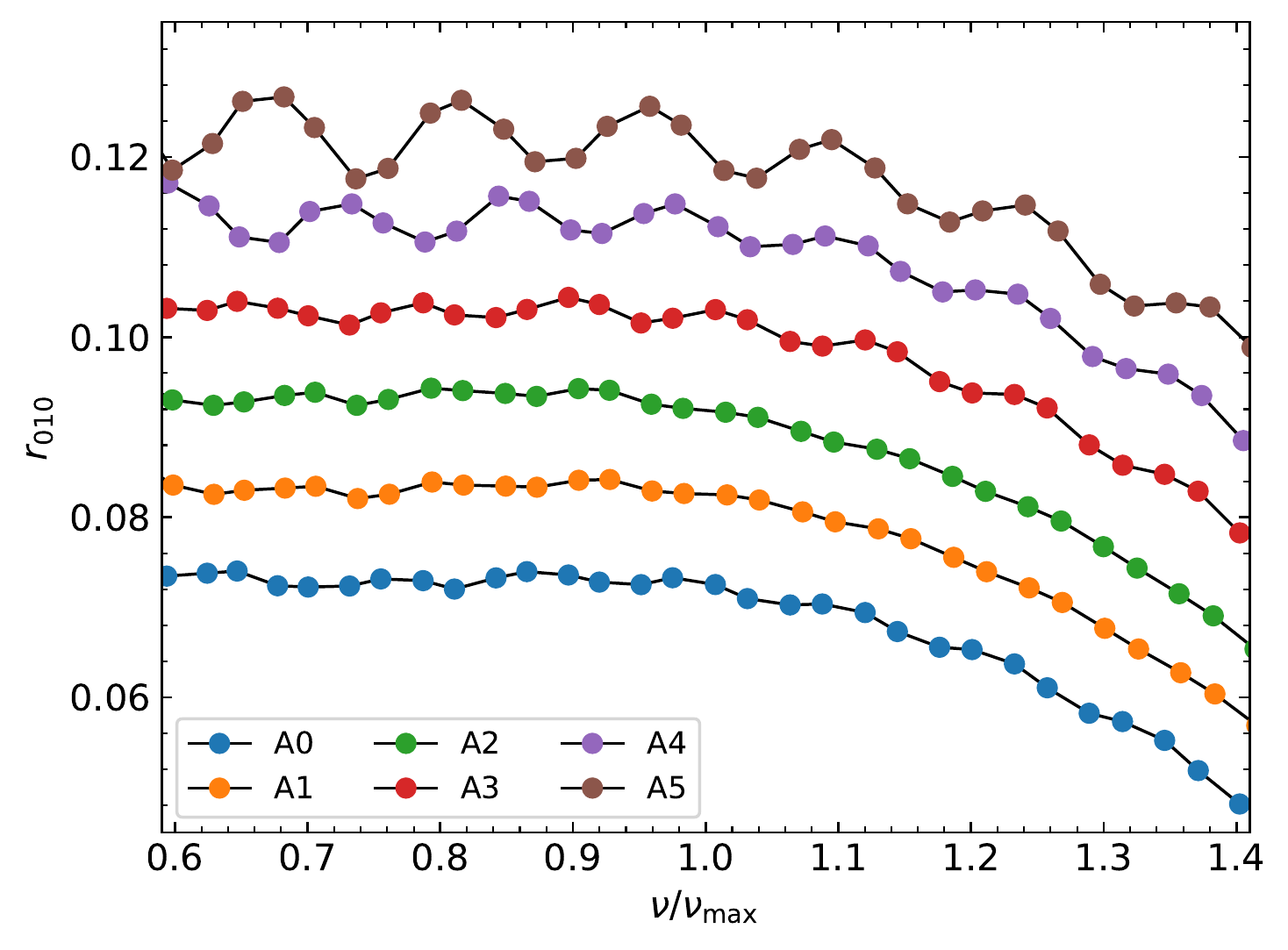}
   \caption{$r_{010}$ ratios for the $1.4 M_\odot$ models A0 to A5 all at $X_c=0.10$. All curves have a mean of about 0.07, and all but A0 are gradually shifted by $0.01$ for clarity.}
   \label{fig:2}
\end{figure}

All stellar models presented in this section are compared at the same evolutionary status, that is, at the moment at which their hydrogen content in the core is $X_c=0.10$. Figure~\ref{fig:2} shows the $r_{010}$ ratios for models A0 to A5. The ratios exhibit a smooth component that is qualitatively similar to that of KIC10162436. The standard A0 model shows no large oscillatory component. The changes in the input physics affecting $\Gamma_1$ and $\mu$ have no significant impact on the ratios (models A1 and A2 in Fig.~\ref{fig:2}). The only scenario inducing a large oscillatory component (i.e. that could be detected despite the larger uncertainties on the frequencies of F-type stars) of those we tested is the addition of a large region of PC (model A5). This conclusion is consistent with that of \cite{lebreton12} and \cite{christophe19}.
We therefore theoretically explore the possibility of large PC regions. The observed F-type stars will be analysed in a future work.

Assuming the large-amplitude signature seen in the ratios $r_{010}$ of F-type stars is linked to large PC regions, the period of the signal is related to $t_\mathrm{cz}$, the acoustic radius of the base of the PC region, and the amplitude is larger when the PC region is larger. This aspect is analytically demonstrated in Section~\ref{glitch_r010} and \ref{detect}. 

\begin{figure*}
    \centering
    \includegraphics[scale=0.6]{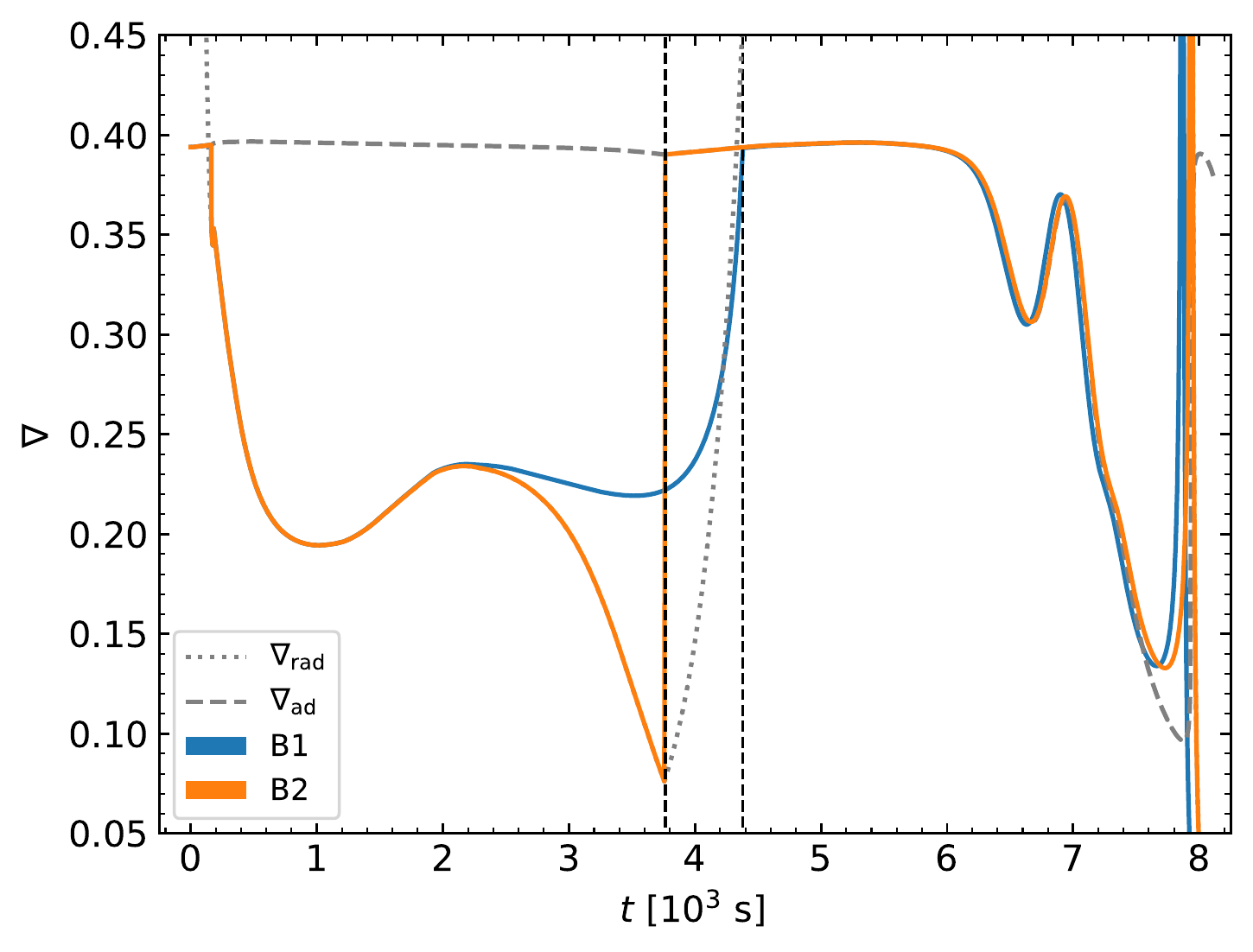}
    \put(-111,170){\large $\big\downarrow$}
    \includegraphics[scale=0.6]{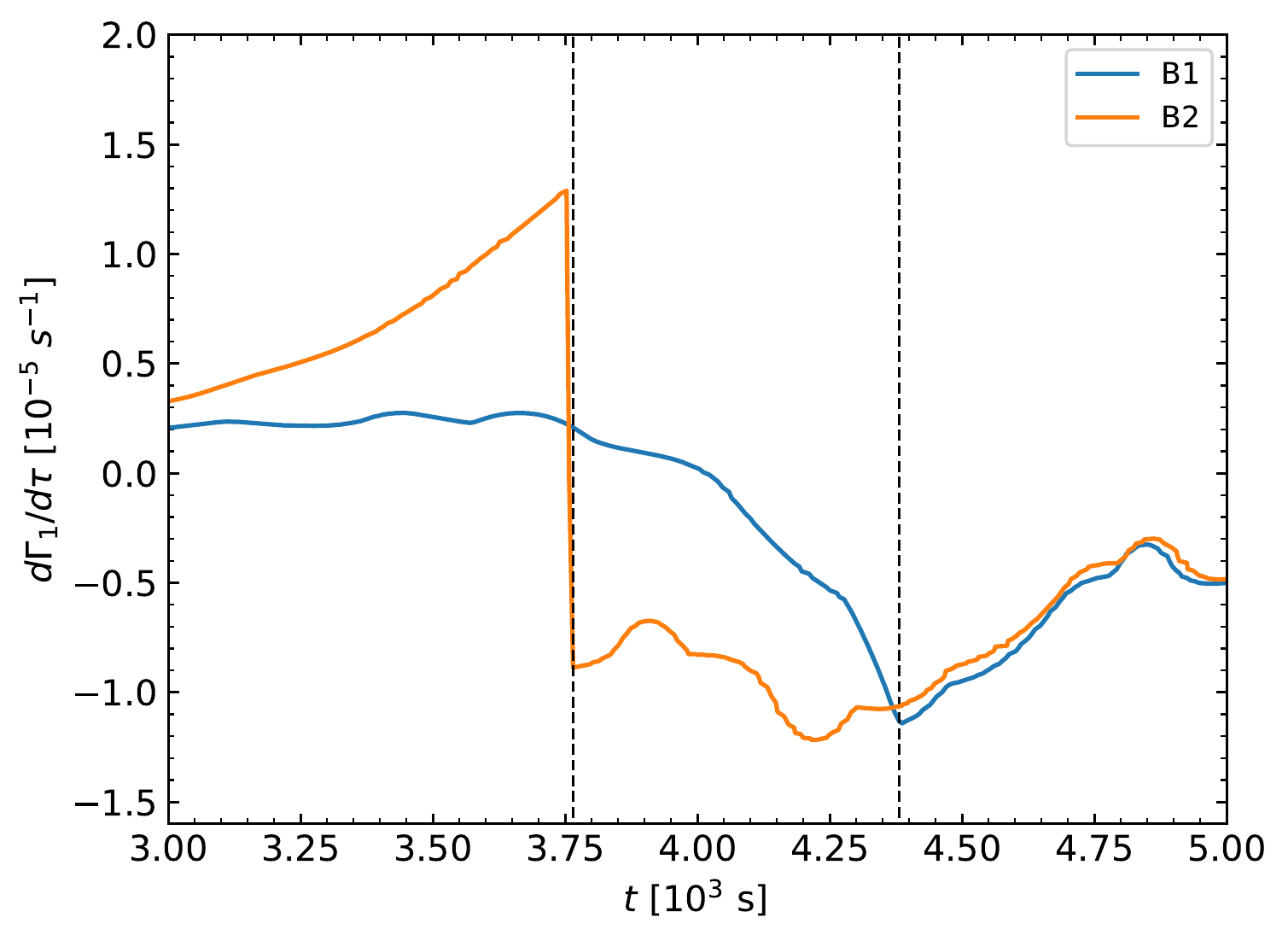}
    \caption{Internal structure profiles with and without penetrative convection. \textbf{Left:} Temperature gradients as function of acoustic radius for the B1 and B2 models with $M=1.4~M_\odot$ and $X_c=0.15$. The grey lines represent the adiabatic and radiative gradients of model B2 (which are very similar to those of B1). The vertical dashed black line on the right represents the position of the Schwarzschild limit for both models. The line on the left represents the bottom of the penetrative convective region of model B2. The downward arrow shows the middle of the acoustic cavity ($t\mathrm{cz}/\mathcal{T}\approx4180~s$). \textbf{Right:} $d\Gamma_1/d\tau$ as a function of the acoustic radius for models B1 and B2.}
   \label{fig:3}
\end{figure*}

\begin{figure}
    \centering
    \includegraphics[scale=0.6]{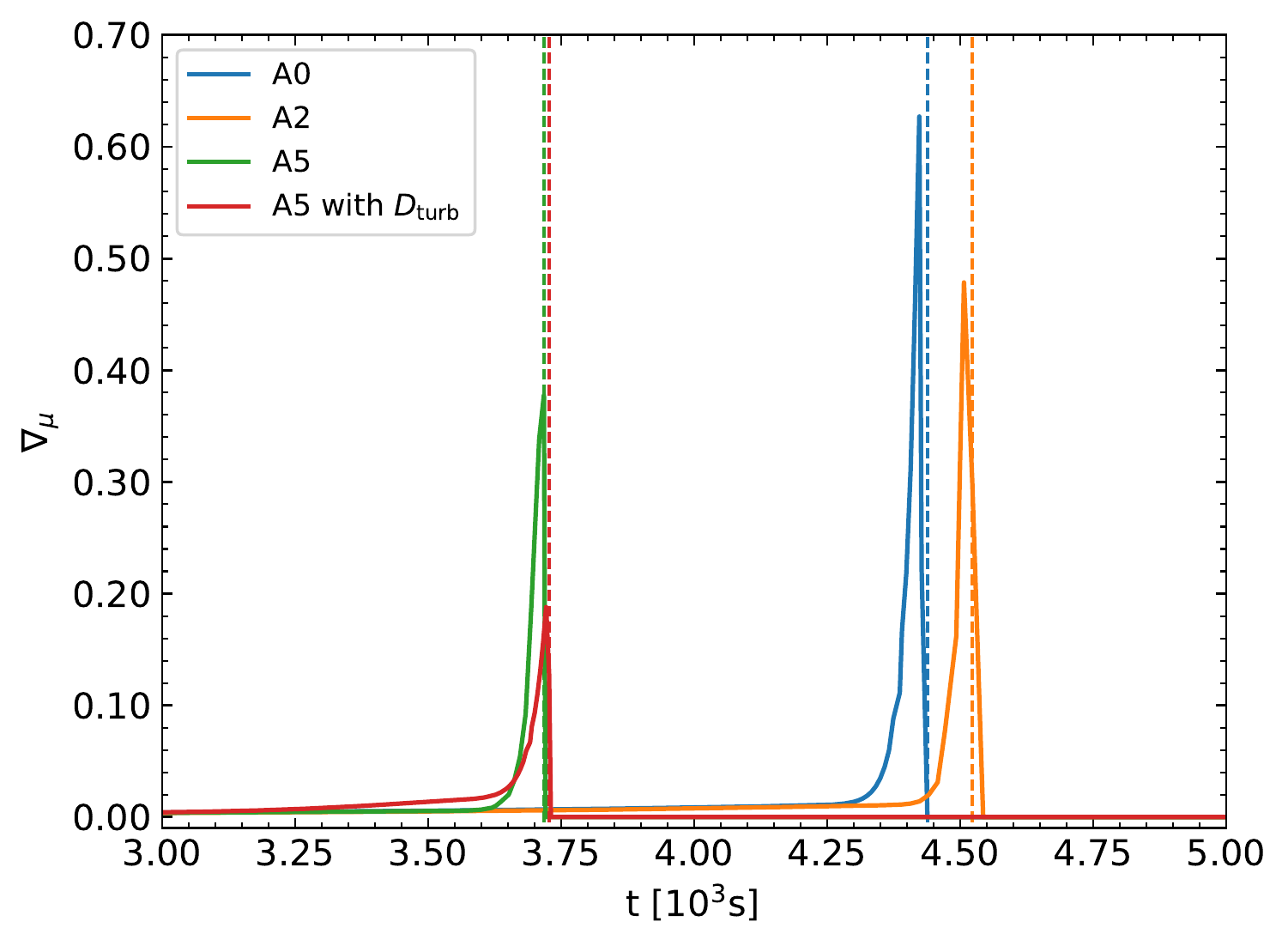}
   \caption{$\mu$ gradient as a function of the acoustic radius for models A0, A5, and A5 with the turbulent mixing calibrated by \cite{verma19b}.}
   \label{fig:3b}
\end{figure}

\section{Theoretical  BSCZ glitch signature}\label{analytics}

In this section, we determine the analytical expression of the amplitude of the signal in the $r_{010}$ ratios according to the structural quantities, similarly to what \cite{monteiro94} and \cite{roxburgh94} reported for the frequency and phase shift variations, respectively. This expression is useful to extract valuable information about the convection to radiation transition in stars for which the amplitude of the glitch signature is larger in the $r_{010}$ ratios than in the frequency variations. We considered the expression for the frequency variation as given by \cite{monteiro94} and propagated it into Eq.~\ref{eq:r010}, assuming  a fully adiabatic PC region \citep{zahn91}. For convenience, specific models without atomic diffusion (B1 and B2) were computed and used only in this section in order to satisfy the assumption on $\mathrm{d}\mu/\mathrm{d}\tau$ (see below). Their input physics are presented at the bottom of Table~\ref{tab:1}.

\begin{figure*}
    \centering
    \includegraphics[scale=0.6]{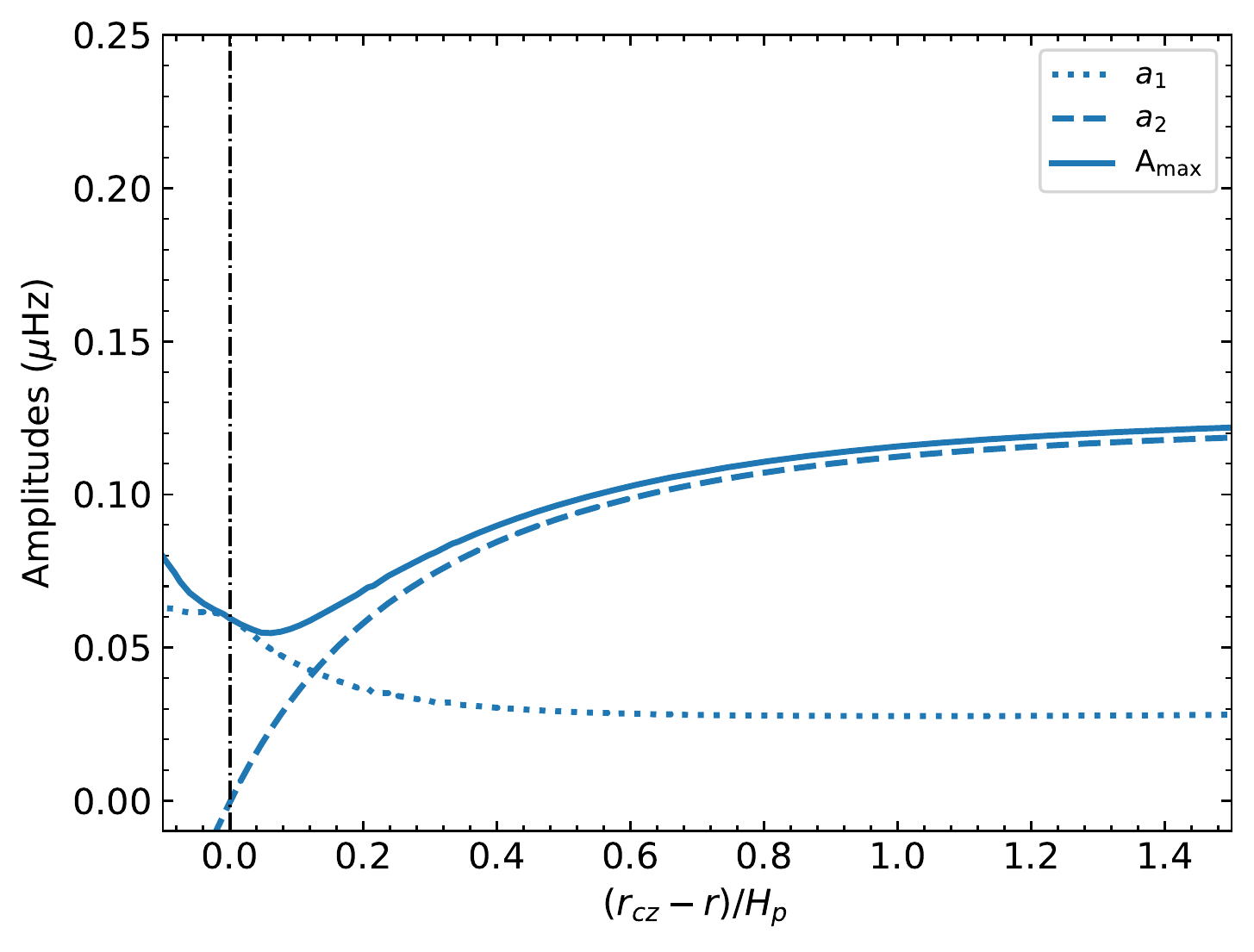}
    \includegraphics[scale=0.6]{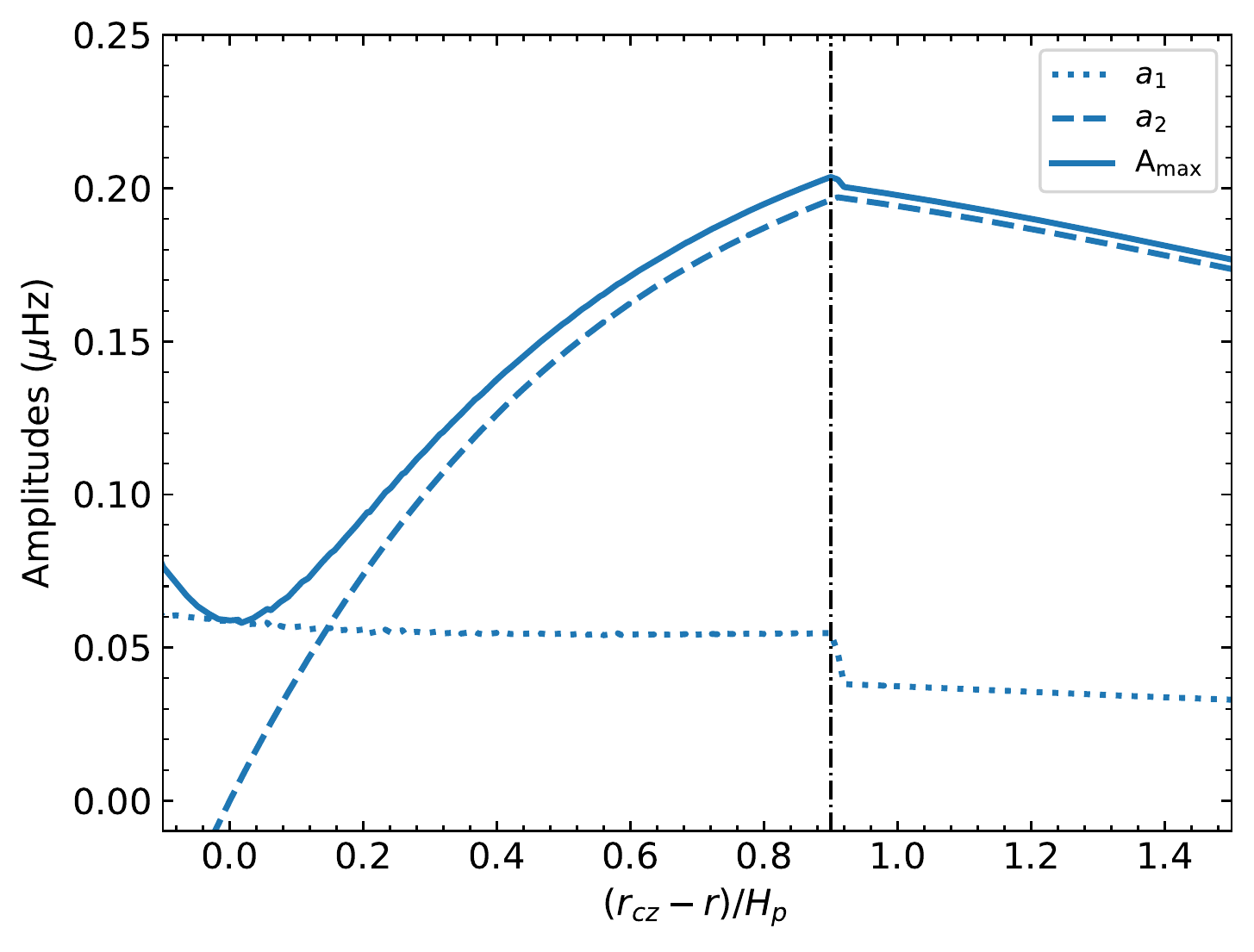}
    \includegraphics[scale=0.6]{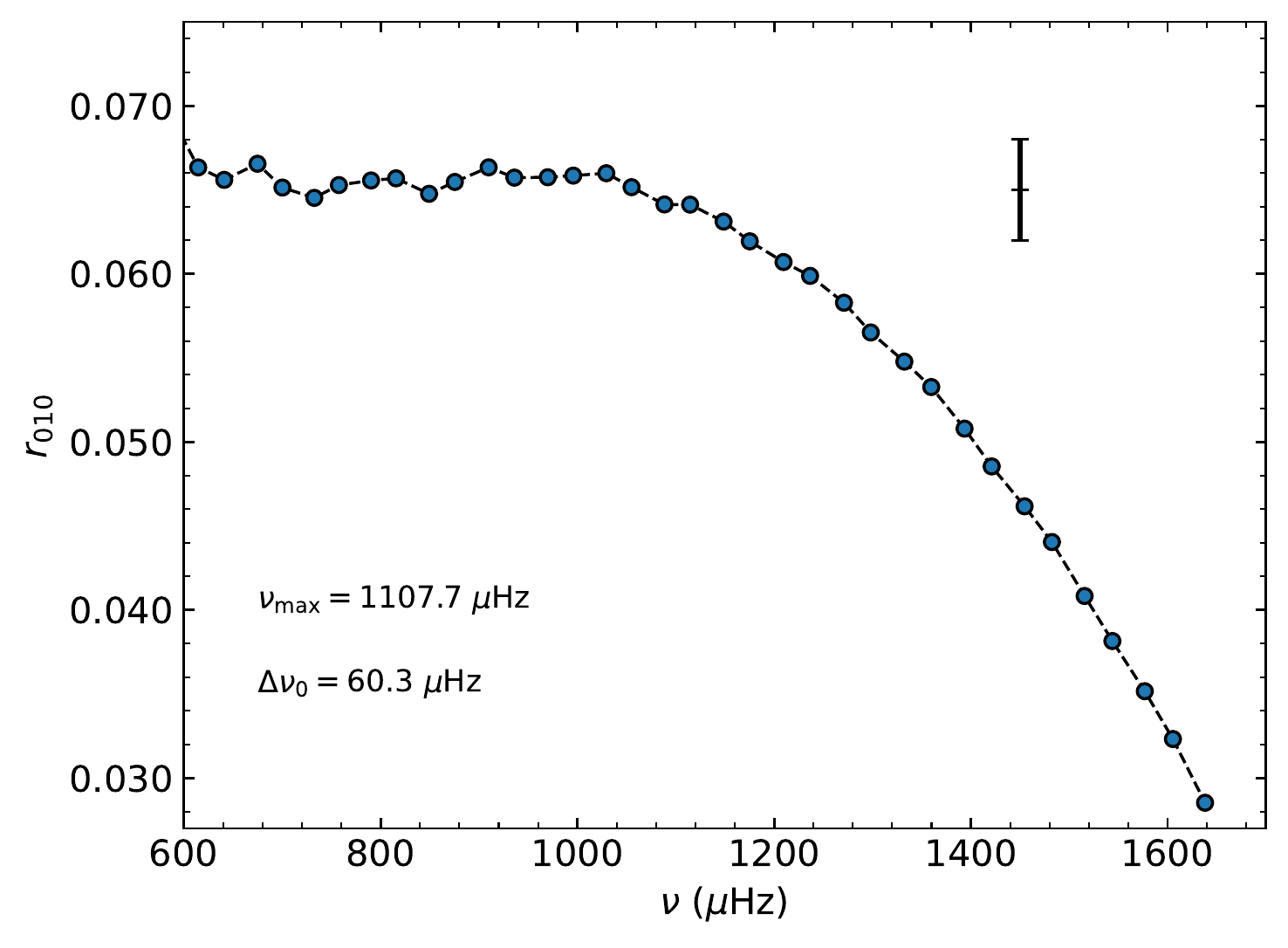}
    \includegraphics[scale=0.6]{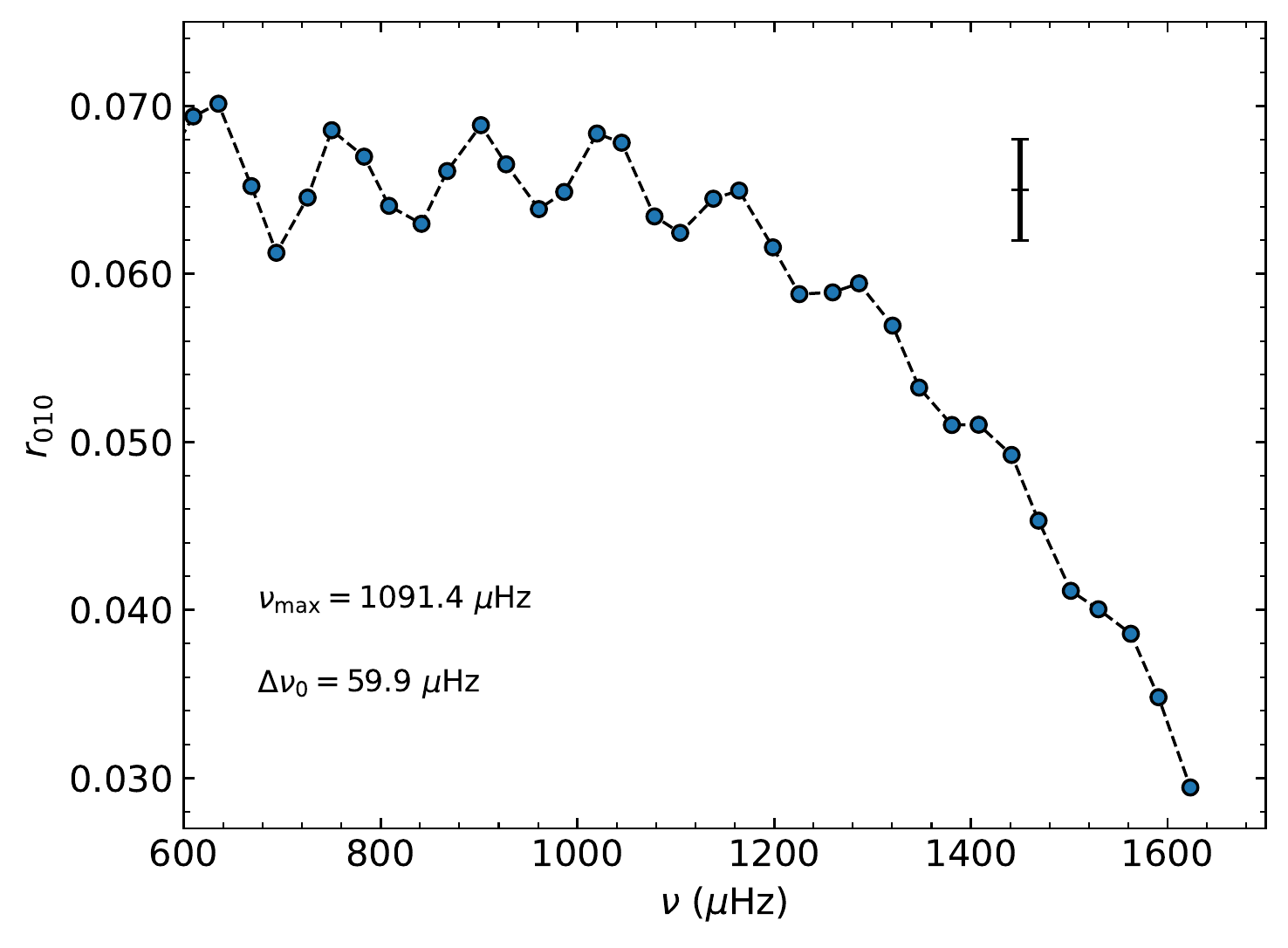}
   \caption{Glitch parameters and signature with and without penetrative convection. \textbf{Top:} Amplitudes $a_1$ (dotted lines), $a_2$ (dashed lines), and $A_\mathrm{max}$ (solid lines) (Eq.~\ref{eq:a1}, \ref{eq:a2}, and \ref{eq:anumax}) as  functions of the extent of the PC region in pressure scale height for models B1 (left panel) and B2 (right panel) at $X_c=0.15$. The dot-dashed lines represent the BSCZ. \textbf{Bottom:} Ratios $r_{010}$ according to the frequency for the same models. The error bar represents the typical mean uncertainty on the ratios, such as the one obtained for  KIC10162436.}
   \label{fig:4}
\end{figure*}
\subsection{Signal in the frequency variations}\label{sect4.1}

The frequency variations induced by the sharp change in temperature gradient from adiabatic to radiative at the BSCZ is described by \citep{monteiro94,roxburgh94}
\begin{equation}\label{eq:mario}
    \begin{aligned}
    \delta\nu_{glitch}(\nu)=&a_1(\tau_d)\left(\frac{\tilde{\nu}}{\nu}\right)^2\sin(4\pi\nu\tau_d+2\phi)\\+&a_2(\tau_d)\left(\frac{\tilde{\nu}}{\nu}\right)\cos(4\pi\nu\tau_d+2\phi),
    \end{aligned}    
\end{equation}
\noindent where $\tilde{\nu}$ is a reference frequency, $\tau_d$ is the acoustic depth of the BSCZ, and $\phi$ is some constant phase.
The above expression can be rewritten as
\begin{equation}\label{eq:mario2}
    \delta\nu_{glitch}(\nu)= A(\nu)\cos(4\pi\nu\tau_d+\phi'(\nu)),
\end{equation}    
\noindent with 
\begin{align}
    A(\nu)=&\left[a_1(\tau_d)^2\left(\frac{\tilde{\nu}}{\nu}\right)^4+a_2(\tau_d)^2\left(\frac{\tilde{\nu}}{\nu}\right)^2\right]^{1/2},\\\label{eq:amp_tot}
    \phi'(\nu)=&2\phi-\arctan\left[\frac{a_1(\tau_d)}{a_2(\tau_d)}\left(\frac{\tilde{\nu}}{\nu}\right)\right].
\end{align}    

The amplitudes $a_1$ and $a_2$ (expressed in Hz) are related to the structure of the star at $\tau_d$ by
\begin{align}
    & a_1(\tau)=\frac{g}{32\pi^3\tilde{\nu}^2\mathcal{T}}\left[h_1(\tau)h_2(\tau)-\frac{1}{4}\left(\frac{\gamma-1}{\nabla_\mathrm {ad}}\frac{\mathrm{d}\nabla_\mathrm {rad}}{\mathrm{d}\tau}\right)\right]\label{eq:a1},\\
    & a_2(\tau)=-\frac{g}{32\pi^2 c_s\tilde{\nu}\mathcal{T}}h_1(\tau)\label{eq:a2},
\end{align}
\noindent with
\begin{align}
    h_1(\tau)&=(\gamma-1)\frac{\nabla_\mathrm{rad}-\nabla_\mathrm{ad}}{\nabla_\mathrm{ad}},\\
    h_2(\tau)&=\frac{g}{c_s}\left(\frac{3h_1(\tau)}{16} - \frac{\Gamma_1+3}{8} + \frac{U-2}{4V_g}\right).
\end{align}

\noindent Here, $U $ and $V_g$ are defined as (\citealt{unno89}, Eq18.18)
$$  U = \frac{4\pi \rho r^3}{m}  =\frac{3 \rho}{\bar \rho} ; V_g = \frac{gr}{c_s^2}.$$

\noindent $\mathcal{T}=t(R^\ast)$ is the total acoustic radius of the star, $g$ is the local gravity, $\gamma$ is the ratio of specific heats and is equal to $\Gamma_1=5/3$ in the perfect gas approximation, $r$ is the local radius, $G$ is the gravitational constant, and  $\nabla_\mathrm{ad}$ and $\nabla_\mathrm{rad}$ are the adiabatic and radiative temperature gradients, respectively. The total amplitude of the signal at $\nu=\tilde{\nu}=\nu_\mathrm{max}$ is  defined by
\begin{equation}\label{eq:anumax}
   A_\mathrm{max}=\left(a_1^2+a_2^2\right)^{1/2}.
\end{equation}

These expressions are obtained under the following assumptions for the region around the BSCZ:
\begin{itemize}
    \item[$\bullet$] $\displaystyle \frac{\nabla-\nabla_\mathrm{ad}}{\nabla_\mathrm{ad}}=\frac{\nabla_\mathrm{rad}-\nabla_\mathrm{ad}}{\nabla_\mathrm{ad}}H(\tau-\tau_d)$, \mbox{with} $H(\tau)$ the Heaviside function. For a fully adiabatic penetrative convection region, this assumption is verified close to the transition region, as shown in the left panels of Figs.~\ref{fig:2bis} and \ref{fig:3}.\\
    \item[$\bullet$] $ \displaystyle \frac{\mathrm{d}\Gamma_1}{\mathrm{d}\tau}\approx 0$ (see right panel of Fig.~\ref{fig:3} for the validity of this assumption, i.e. $d\Gamma_1/d\tau<1.5\times10^{-5}~s^{-1}$)\\
    \item[$\bullet$]  $ \displaystyle\frac{\mathrm{d}\mu}{\mathrm{d}\tau}\approx 0$,\\
    \item[$\bullet$] the region is fully ionised, and the pressure in the fully ionised region is approximated by ideal gas plus radiative pressure.
\end{itemize}
In the context of a fully adiabatic PC region, all these assumptions are reasonable, except for the neglect of a possible gradient in the mean molecular weight that can arise because of the effect of atomic diffusion (see Fig.~\ref{fig:3b} and model A2 in Fig.~\ref{fig:2}). Accordingly, in this section, our stellar models are computed assuming no atomic diffusion in order to avoid any potential contamination of the signal from any sharp variations in $\mu$ in the relevant region. The variation in the amplitude induced by atomic diffusion (i.e. the $\mu$ gradient) has already been discussed for the Sun \citep{basu94a}. The impact of this specific assumption for F-type stars will be estimated in a future work.

Figure~\ref{fig:4} shows $a_1$, $a_2$ , and $A_\mathrm{max}$ for models B1 and B2 (see Table~\ref{tab:A0} and \ref{tab:1} for the input physics). Below the Schwarzschild convective zone limit ($(r_\mathrm{cz}-r)/H_p>0.1$), $a_1$ decreases and $a_2$ increases. $a_1$ is only slightly affected by the inclusion of a PC (model B2) region, whereas $a_2$ becomes significantly larger (by about a factor of 2) than in absence of convective penetration, and it therefore dominates the $a_1$ contribution in the presence of PC. Moreover, the deeper the PC region, the higher the $a_2$ value because the difference between the adiabatic and radiative temperature gradients increases.

\subsection{Signal in the $r_{010}$ ratios}\label{glitch_r010}

\begin{figure}
    \centering
    \includegraphics[scale=0.6]{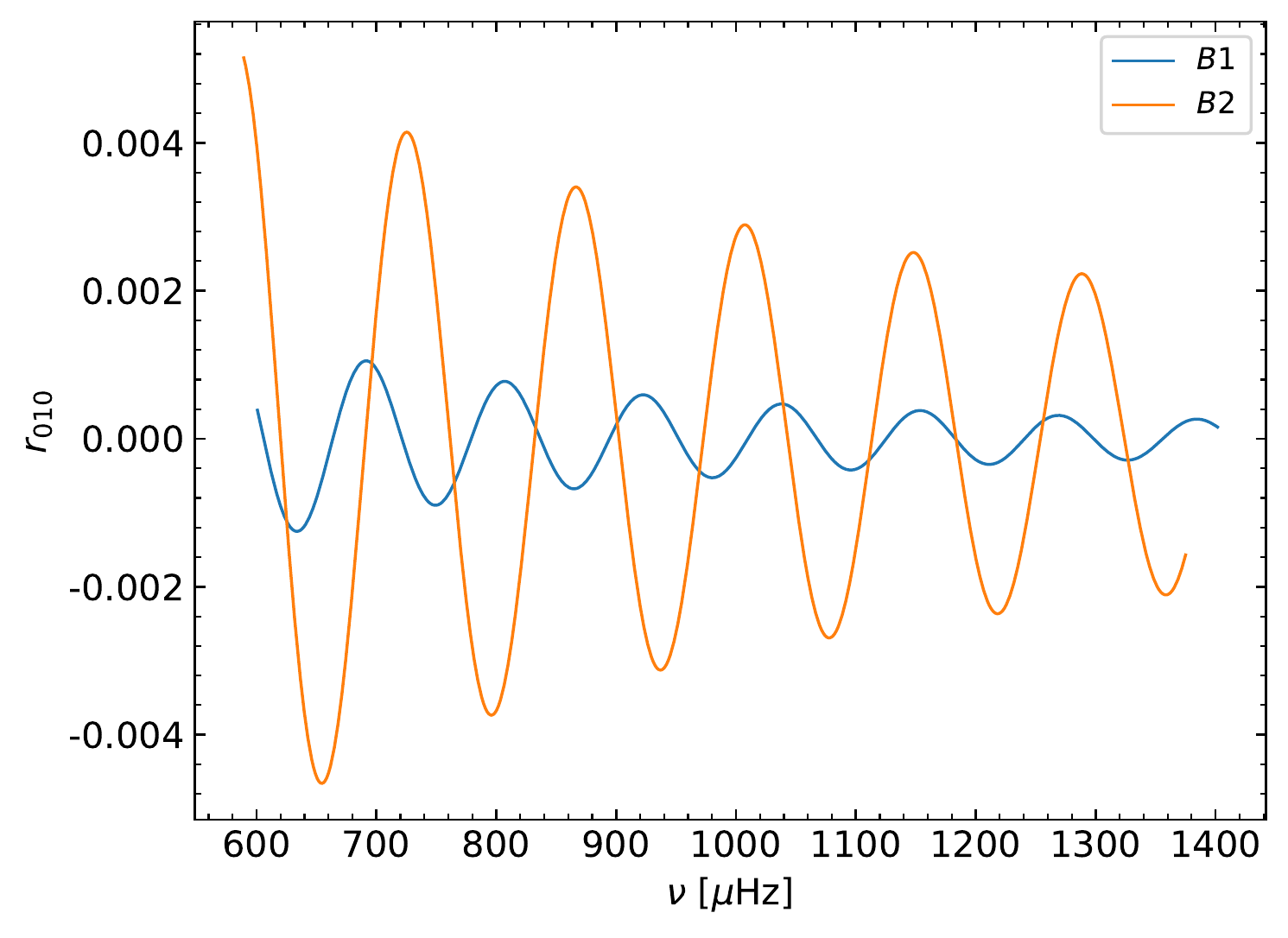}
    \caption{Predicted glitch signature from Eq.~\ref{eq:r010-th} for model B1 and B2 at $X_c=0.15$.}
   \label{fig:5}
\end{figure}

We now use the expression of the frequency variation (Eq.~\ref{eq:mario}) and inject it in Eq.~\ref{eq:r010}. We obtain the following expression for the ratio of small to large separations of $l=0,1$ models (see Appendix~\ref{appendix:A} for the detailed calculation):
\begin{equation}\label{eq:r010-th}
    \begin{aligned}
        r_{010,~\mathrm{glitch}}(\nu)\approx& a_1(\tau_d)\left(\frac{\tilde{\nu}}{\nu}\right)^2  \times \frac{1}{4\bar{\Delta}} f_{12}(\nu)\times\sin\left(4\pi\nu t_d+2\phi\right)\\
        +&a_2(\tau_d) 
      \left(\frac{\tilde{\nu}}{\nu}\right)\times   \frac{1}{4\bar{\Delta}} f_{21}(\nu)\times\cos\left(4\pi\nu t_d+2\phi\right),
    \end{aligned}
\end{equation}
with
\begin{equation}
    \begin{aligned}
        f_{12}(\nu)=&3+\left[\frac{\nu^2(\nu^2+\bar{\Delta}^2)}{(\nu^2-\bar{\Delta}^2)^2}\right]\cos\left(4\pi\bar{\Delta}t_d\right)\\
        +& \left[\frac{4\nu^2(\nu^2+\bar{\Delta}^2/4)}{(\nu^2-\bar{\Delta}^2/4)^2}\right]\cos\left(2\pi\bar{\Delta}t_d\right)\label{eq:f12},
    \end{aligned}
\end{equation}

\begin{equation}
    \begin{aligned}
        f_{21}(\nu)=&3+ \left[\frac{\nu^2}{\nu^2 - \bar{\Delta}^2}\right]\cos\left(4\pi\bar{\Delta}t_d\right) \\
        +& \left[\frac{4\nu^2}{\nu^2-\bar{\Delta}^2/4}\right]\cos\left(2\pi\bar{\Delta}t_d\right)\label{eq:f21},
    \end{aligned}
\end{equation}
\noindent where $\bar \Delta$ is the mean large separation assumed to be the same for the $l=0$ and $l=1$ mode degrees.

Because the period of the signature of the BSCZ in the $r_{010}$ ratios is the acoustic radius of the BSCZ rather than its acoustic depth \citep{roxburgh09}, we converted the expressions so that they are written in terms of $t_d$ instead of $\tau_d$.

Because $\bar \Delta /\nu <<1$,
\begin{eqnarray}
f_{12}(\nu)\approx f_{21}(\nu) \approx \Bigl( \cos\left(2 \Phi_t  \right) + 4\cos\left(\Phi_t  \right)+3 \Bigr)    \leq  8 ,      
\end{eqnarray}
where $\Phi_t= 2 \pi \bar \Delta t_d $. We then recover the same frequency dependence as  for $\delta \nu$ in Eq.\ref{eq:mario}, except for an amplification factor $\leq 2/\bar \Delta$. For model B2 at $X_c=0.15$, the amplitude of $\delta\nu$ is $\leq 0.2~\mu$Hz and that of $r_{010,~\mathrm{glitch}}$ is $\leq 0.0067$. This is to be compared with the uncertainties around $\nu_\mathrm{max}$ of KIC10162436, for instance (Fig.~\ref{fig:1}), which are $0.17~\mu$Hz and $0.0030$ for the frequencies and the ratios, respectively. Whereas the maximum  amplitude of $\delta\nu$  is very close to the uncertainty, the amplitude of the signature in the ratios is amplified up to more than a factor of 2, facilitating the signature detection with the ratios.

For $\tilde{\nu}=\nu_\mathrm{max}$ and  for frequencies $\nu$ around $\nu_\mathrm{max}$, the expression can be written
\begin{equation}\label{eq:r010_numax}
        r_{010,~\mathrm{glitch}}(\nu)\approx\frac{A_\mathrm{max}}{4\bar{\Delta}}\times f_{12/21}(\nu_\mathrm{max})\times\sin\left(4\pi\nu t_d+\phi'\right),
\end{equation}
\noindent where $f_{12/21}(\nu_\mathrm{max})\approx f_{12}(\nu_\mathrm{max})\approx f_{21}(\nu_\mathrm{max})$. In this case, the ratios approximately depend on a constant amplitude.

\section{Validation}\label{valid}
\subsection{Validation of the fitting procedure}
\begin{table*}[!ht]
\centering
\caption{Cases considered to validate the fitting procedure with model B2.} 
\label{tab:3}
\begin{tabular}{ccccc c ccccc}
\hline\hline
Case & $X_c$ & $\tilde{\nu}$ [$\mu$Hz] & $\bar{\Delta}$ [$\mu$Hz] & $\bar{\sigma_\mathrm{freq}}$ [$\mu$Hz] & Fitting eq. & $a_1$ [$\mu$Hz] & $a_2$ [$\mu$Hz] & $A_\mathrm{max}$ [$\mu$Hz] & $t_d$ [s] & $\phi$ [radian] \\
\hline
%
  & $0.25$  & $1232.6$ & $65.1$ & - & Theoretical & $0.056$ & $0.225$ & $0.232$ & $3931$ & - \\
\cmidrule(lr){5-11}
\smallskip
1 &  &  & & $0.10$ & Eq.~\ref{fit:2} & $0.100^{+0.068}_{-0.067}$ & $0.185^{+0.041}_{-0.099}$ & $0.220^{+0.026}_{-0.032}$ & $3943^{+32}_{-37}$  & $0.83^{+0.25}_{-0.27}$ \\
\smallskip
&  &  & & $0.10$ & Eq.~\ref{fit:3} & - & $0.227^{+0.026}_{-0.028}$ & $0.227^{+0.026}_{-0.028}$ & $3957^{+27}_{-34}$ & $0.91^{+0.46}_{-0.37}$ \\
\cmidrule(lr){5-11}
\smallskip
2 &  &  & & $0.17$ & Eq.~\ref{fit:2} & $0.132^{+0.066}_{-0.081}$ & $0.145^{+0.077}_{-0.095}$ & $0.212^{+0.041}_{-0.043}$ & $3830^{+83}_{-81}$  & $1.86^{+0.54}_{-0.61}$ \\
\smallskip
&  &  & & $0.17$ & Eq.~\ref{fit:3} & - & $0.218^{+0.043}_{-0.046}$ & $0.218^{+0.043}_{-0.046}$ & $3844^{+83}_{-79}$ & $1.41^{+0.60}_{-0.51}$ \\
\hline
%
 & $0.15$  & $1091.4$ & $59.8$ & - & Theoretical & $0.055$ & $0.196$ & $0.204$ & $3757$ &- \\
\cmidrule(lr){5-11}
\smallskip
3 &  &  & & $0.10$ & Eq.~\ref{fit:2} & $0.160^{+0.026}_{-0.044}$ & $0.101^{+0.068}_{-0.065}$ & $0.194^{+0.020}_{-0.018}$ & $3716^{+38}_{-38}$  & $1.54^{+0.25}_{-0.28}$ \\
\smallskip
&  &  & & $0.10$ & Eq.~\ref{fit:3} & - & $0.207^{+0.023}_{-0.030}$ & $0.207^{+0.023}_{-0.030}$ & $3736^{+37}_{-41}$  & $0.913^{+23}_{-0.24}$ \\
\cmidrule(lr){5-11}
\smallskip
4 &  &  & & $0.17$ & Eq.~\ref{fit:2} & $0.102^{+0.068}_{-0.068}$ & $0.120^{+0.077}_{-0.075}$ & $0.179^{+0.052}_{-0.056}$ & $3751^{+125}_{-157}$  & $1.25^{+1.014}_{-0.728}$ \\
\smallskip
&  &  & & $0.17$ & Eq.~\ref{fit:3} & - & $0.188^{+0.045}_{-0.045}$ & $0.188^{+0.045}_{-0.045}$ & $3770^{+120}_{-136}$  & $1.07^{+1.4}_{-0.69}$ \\
\cmidrule(lr){5-11}
\smallskip
5 &  &  & & $0.30$ & Eq.~\ref{fit:2} & $0.098^{+0.094}_{-0.068}$ & $0.108^{+0.100}_{-0.073}$ & $0.180^{+0.083}_{-0.077}$ & $3751^{+246}_{-292}$  & $1.48^{+1.21}_{-1.01}$ \\
\smallskip
&  &  & & $0.30$ & Eq.~\ref{fit:3} & - & $0.143^{+0.097}_{-0.090}$ & $0.143^{+0.097}_{-0.090}$ & $3753^{+321}_{-355}$  & $1.55^{+1.09}_{-1.07}$ \\
%
\hline
 & $0.05$  & $982.2$ & $55.5$ & - & Theoretical & $0.052$ & $0.183$ & $0.190$ & $3575$  & -  \\
\cmidrule(lr){5-11}
\smallskip
6 &  &  & & $0.10$ & Eq.~\ref{fit:2} & $0.058^{+0.049}_{-0.040}$ & $0.149^{+0.031}_{-0.050}$ & $0.163^{+0.024}_{-0.024}$ & $3535^{+52}_{-56}$  & $1.27^{+0.40}_{-0.36}$ \\
\smallskip
&  &  & & $0.10$ & Eq.~\ref{fit:3} & - & $0.174^{+0.022}_{-0.023}$ & $0.174^{+0.022}_{-0.023}$ & $3563^{+47}_{-51}$ & $0.89^{+0.30}_{-0.26}$ \\
\cmidrule(lr){5-11}
\smallskip
7 &  &  & & $0.17$ & Eq.~\ref{fit:2} & $0.116^{+0.060}_{-0.071}$ & $0.119^{+0.075}_{-0.080}$ & $0.183^{+0.045}_{-0.044}$ & $3490^{+140}_{-118}$  & $1.71^{+0.65}_{-0.73}$ \\
\smallskip
&  &  & & $0.17$ & Eq.~\ref{fit:3} & - & $0.181^{+0.054}_{-0.051}$ & $0.181^{+0.054}_{-0.051}$ & $3521^{+148}_{-131}$ & $1.26^{+0.79}_{-0.69}$ \\
\hline
\end{tabular}
\end{table*} 

\begin{table*}[!ht]
\centering
\caption{Fitted parameters of the ratios for the Sun} 
\label{tab:4}
\begin{tabular}{cc c ccccc}
\hline\hline
$\tilde{\nu}$ [$\mu$Hz] & $\bar{\Delta}$ [$\mu$Hz] & Fitting eq. & $a_1$ [$\mu$Hz] & $a_2$ [$\mu$Hz] & $A_{\tilde{\nu}}$ [$\mu$Hz] & $t_d$ [s] & $\phi$ [radian] \\
\hline
%
\smallskip
$2500$\tablefoottext{a} & $135.1$\tablefoottext{b} & - & - & - & $0.085$\tablefoottext{a} & $1370$\tablefoottext{a} & -\\
- & - & - & - & - & - & $1370\pm110$\tablefoottext{c} & -\\
- & -  & - & - & - & $[0.033;0.089]$\tablefoottext{d} & $[1488;1533]$\tablefoottext{d} & -\\

- & - & - & - & - & - & $1422\pm20$\tablefoottext{e} & -\\
\cmidrule(lr){3-8}
 \smallskip
  & & Eq.~\ref{fit:2} & $0.071^{+0.008}_{-0.015}$ & $0.034^{+0.023}_{-0.022}$ & $0.079^{+0.004}_{-0.004}$ & $1427^{+12}_{-12}$  & $0.39^{+0.19}_{-0.18}$ \\
\smallskip
 & & Eq.~\ref{fit:3} & - & $0.083^{+0.004}_{-0.004}$ & $0.083^{+0.004}_{-0.004}$ & $1434^{+10}_{-10}$  & $2.85^{+0.15}_{-0.17}$ \\
\hline
\end{tabular}
\tablefoot{\tablefoottext{a}{\cite{monteiro94}}, \tablefoottext{b}{\cite{huber11}}, \tablefoottext{c}{\cite{ballot04}}, \tablefoottext{d}{\cite{JCD11}}, \tablefoottext{e}{\cite{roxburgh09}}}
\end{table*} 

In this section, we test the analytical expression of the $r_{010}$ ratios defined in Sect.~\ref{glitch_r010} using stellar models. Oscillation frequencies were computed with the code ADIPLS \citep{christensen08}. The ratios can be divided into two components (smooth and oscillatory) following an expression of the form
\begin{equation}
    \begin{aligned}
     r_{010}(\nu) =& r_{010,~\mathrm{smooth}}(\nu) + r_{010,~\mathrm{glitch}}(\nu).  
    \end{aligned}
\end{equation}

\cite{deheuvels16} (hereafter D16) developed a method for characterising the signal of the smooth component with a second-order polynomial fitting and constrained the extension of convective cores for eight \textit{Kepler} stars. The second-order polynomial expression of the frequency is defined by
\begin{equation}\label{eq:smoothD16}
    P_{\mathrm{D16}}(\nu)=c_0+c_1(\nu-\beta)+c_2(\nu-\gamma_1)(\nu-\gamma_2),
\end{equation}
\noindent where $\beta$, $\gamma_1$ , and $\gamma_2$ should be adapted so that $c_0$, $c_1$ , and $c_2$ are uncorrelated (see Appendix B of D16). For the purpose of this work, the $c$ coefficients do not necessary need to be uncorrelated. We then characterise the smooth component with a simpler form of the second-order polynomial such as  
\begin{equation}\label{eq:smooth}
    P(\nu)=c_0+c_1\nu+c_2\nu^2.
\end{equation}
This expression is used to characterise the smooth component of the ratios in the following sections. The ratios are then fitted using the following expression:

\begin{equation}\label{fit:2}
\begin{aligned}
    r_{010}(\nu)=& P(\nu) + a_1(\tau_d)~\Bigl(\frac{\tilde{\nu}}{\nu}\Bigr)^2
    ~ \times\frac{1}{4\bar{\Delta}} f_{12}(\nu)\times\sin\left(4\pi\nu t_d+2\phi\right)\\
        &+a_2(\tau_d)\Bigl(\frac{\tilde{\nu}}{\nu}\Bigr) ~\times 
        \frac{1}{4\bar{\Delta}} f_{21}(\nu)\times\cos\left(4\pi\nu t_d+2\phi\right),
\end{aligned}
\end{equation}

\noindent where $a_1$, $a_2$, $t_{d}$ , and $\phi$ are the parameters fitted simultaneously with the coefficients $c_0$, $c_1$ , and $c_2$ of the smooth component $P(\nu)$. The reference frequency is defined as $\tilde{\nu}=\nu_\mathrm{max}$. We also tested a second approximated expression,
\begin{equation}
\begin{aligned}\label{fit:3}
    r_{010}(\nu)=& P(\nu) + a_2(\tau_d)~\Bigl(\frac{\tilde{\nu}}{\nu}\Bigr) ~\times\frac{1}{4\bar{\Delta}} f_{21}(\nu)\times\cos\left(4\pi\nu t_d+2\phi\right),
\end{aligned}
\end{equation}
\noindent because $a_2$ dominates $a_1$ in presence of PC. We used model B2 as an illustration. 

We assessed three scenarios:
\begin{itemize}
\item one scenario using precise frequencies with a mean uncertainty of $\sigma_\mathrm{freq}=0.10~\mu$Hz and a frequency range of [0.6; 1.4]~$\nu_\mathrm{max}$ (representative of the total frequency range of KIC10162436),
\item a second scenario considering a mean uncertainty of $0.17~\mu$Hz (representative of KIC10162436) and a frequency range of [0.7; 1.2]~$\nu_\mathrm{max}$ (more centred around $\nu_\mathrm{max}$), 
\item finally, a third scenario considering a mean uncertainty of $0.30~\mu$Hz and a frequency range of [0.7; 1.2]~$\nu_\mathrm{max}$.
\end{itemize}

We finally selected seven test cases. Cases 1 and 2 represent model B2 at $X_c=0.25$ and $\bar{\sigma_\mathrm{freq}}=0.10$ and $0.17~\mu$Hz, respectively. Cases 3, 4, and 5 represent model B2 at $X_c=0.15$ and $\bar{\sigma_\mathrm{freq}}=0.10$, $0.17,$ and $0.30~\mu$Hz, respectively. Finally, cases 6 and 7 represent model B2 at $X_c=0.05$ and $\bar{\sigma_\mathrm{freq}}=0.10$ and $0.17~\mu$Hz, respectively. All the cases are presented in Table~\ref{tab:3}. The uncertainties were generated randomly with a Gaussian distribution around the mean value using a standard deviation of 20\% of the mean value.

The fitting procedure is based on the {\sc{emcee}} Python package \citep{emcee13}. We used 5000 burning steps and 500 production steps for the Markov chain Monte Carlo method (MCMC). Uniform uninformative priors were chosen for $a_1$, $a_2$ ([0.0; 0.4]), $t_\mathrm{cz}$ ([500; 8500] s) and $\phi$ ([-$2\pi$; $4\pi$]) to keep the parameters in the range expected for stellar interiors. Similarly to D16, the covariance matrix $\mathcal{C}$ was estimated with a Monte Carlo simulation using the frequencies of the model and the random uncertainties described above for all cases. Because the covariance matrix is nearly non-invertible, we truncated it using the singular value decomposition (SVD) approach. The likelihood function $\mathcal{L}$ is then
\begin{equation}\label{likelihood}
         \mathcal{L}(D,\Theta)\propto-\frac{1}{2}\left(r_{010,\mathrm{target}}-r_{010,\mathrm{mod}}\right)^T\mathcal{C}^{-1}\left(r_{010,\mathrm{target}}-r_{010,\mathrm{mod}}\right),
\end{equation}
\noindent where $D$ and $\Theta$ represent the target and the model parameters, respectively. The quantity $r_{010,\mathrm{target}}$ represents the ratios of the target models and in later section those of a real star. The quantity $r_{010,\mathrm{mod}}$ represents the ratios obtained with Eq.~\ref{fit:2} and \ref{fit:3}. The results of the fit for each case are presented in Table~\ref{tab:3}.
\begin{figure}
    \centering
    \includegraphics[scale=0.55]{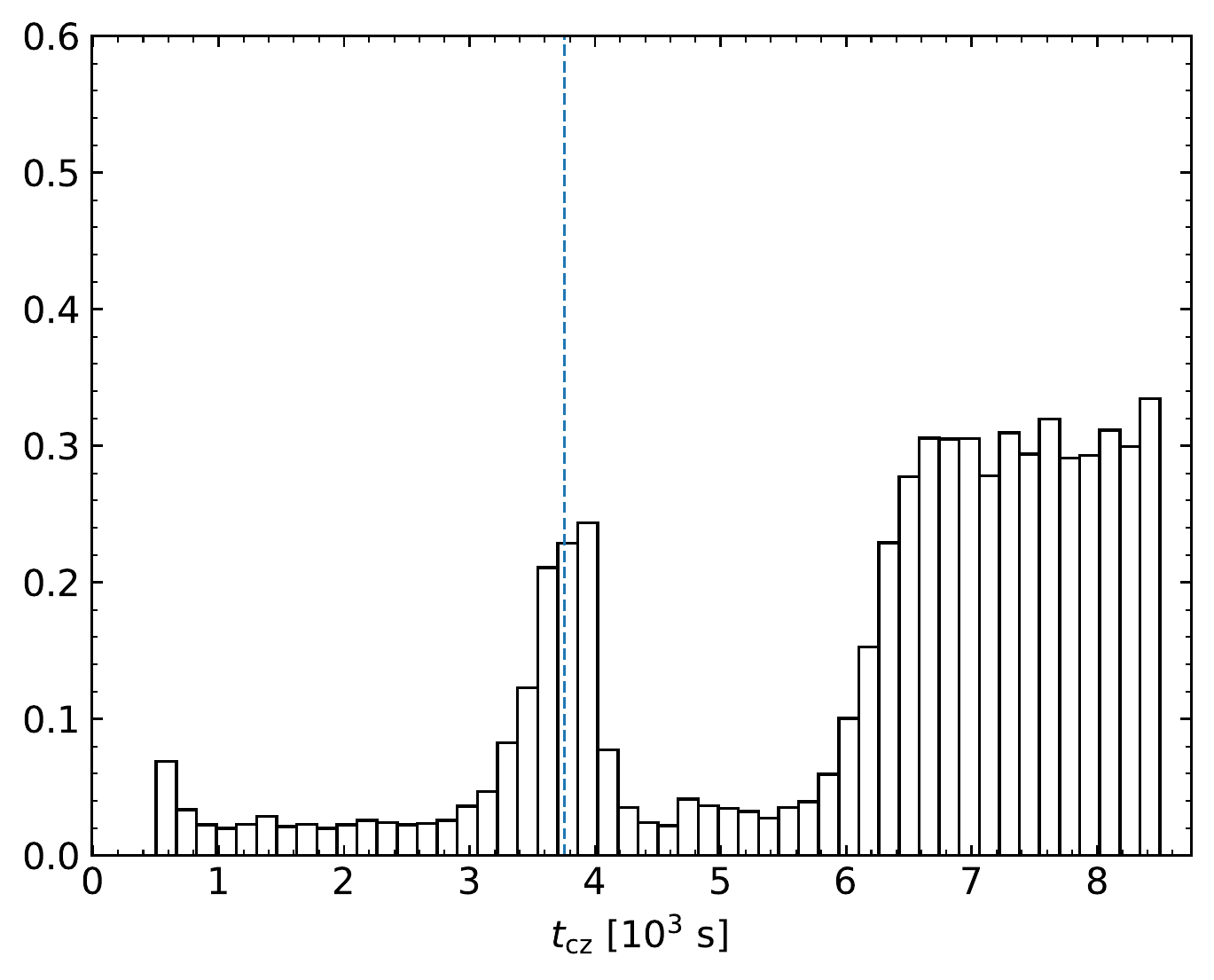}
    \put(-190,150){\tiny $X_c=0.15$, $\bar{\sigma_\mathrm{freq}}=0.30~\mu$Hz}
    \put(-110,100){\large $\big\downarrow$}
   \caption{Probability density of $t_\mathrm{cz}$ when no prior is imposed for the model B2 at $X_c=0.15$ with a mean uncertainty of $0.30~\mu$Hz. The vertical dashed blue lines represent the theoretical value given by the stellar model from which the frequencies were computed. The downward arrow shows the middle of the acoustic cavity ($t_\mathrm{cz}\approx4180~s$).}
  \label{fig:6}
  \end{figure}

\begin{figure*}
    \centering
    \includegraphics[scale=0.61]{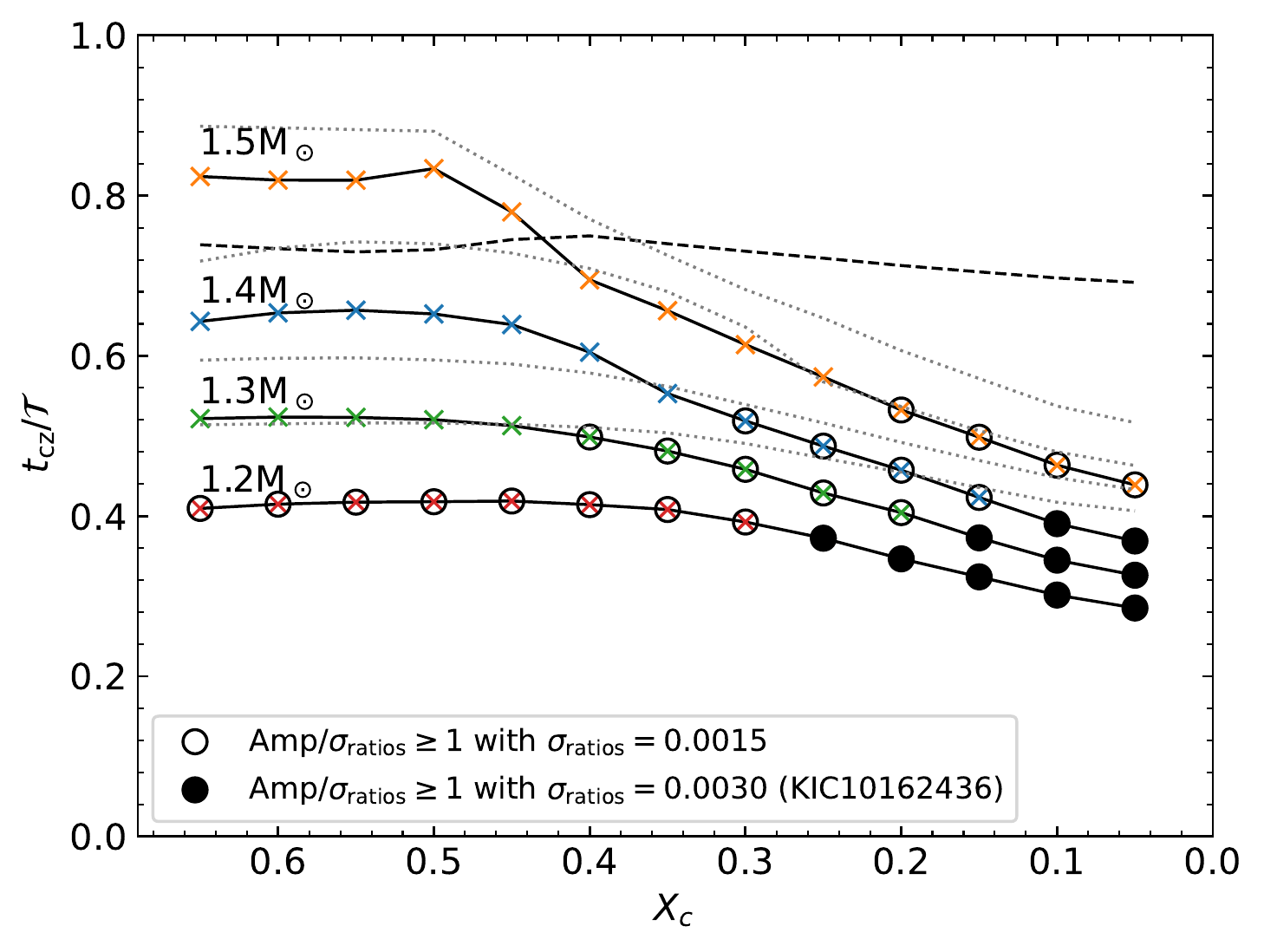}
    \includegraphics[scale=0.61]{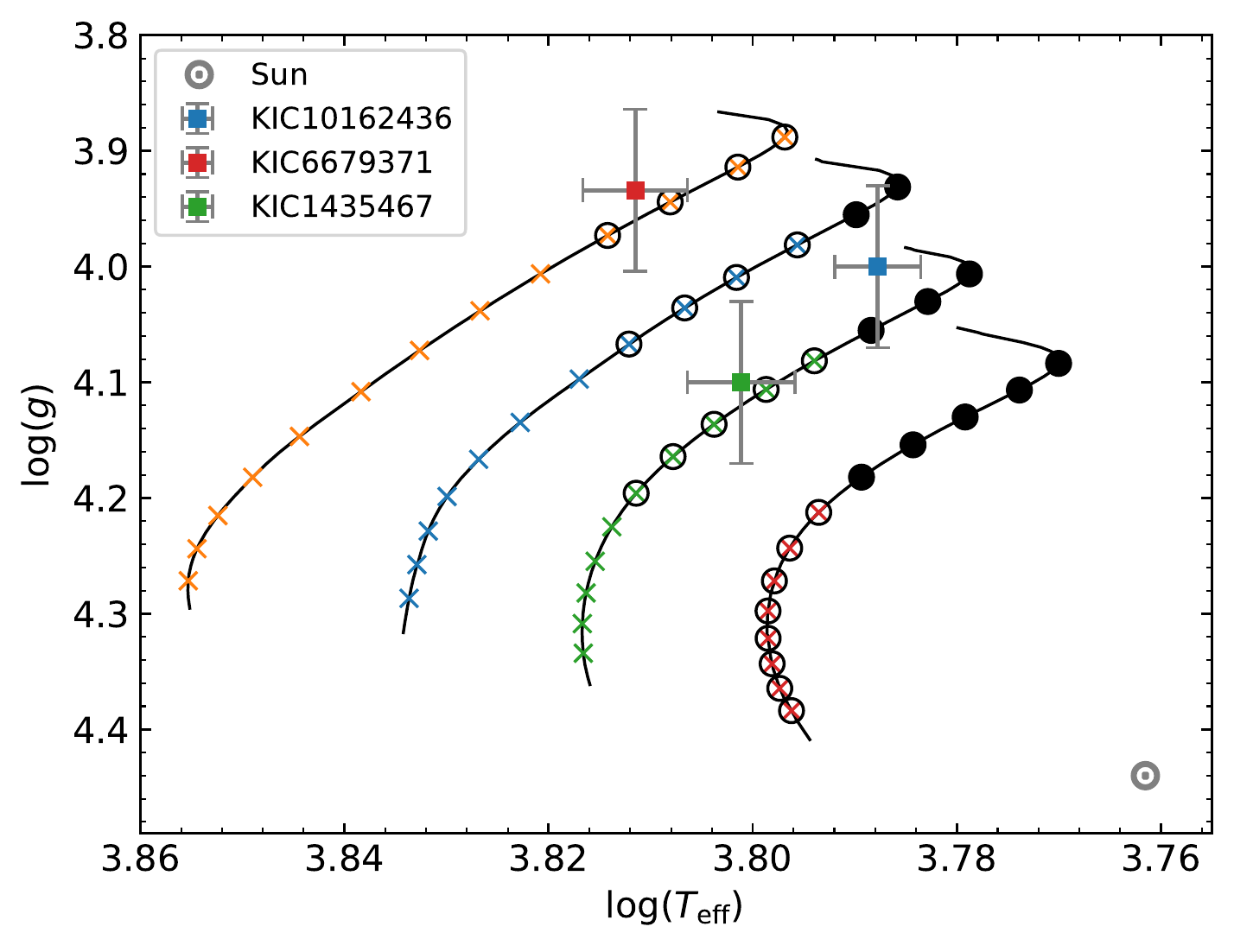}
    \includegraphics[scale=0.6]{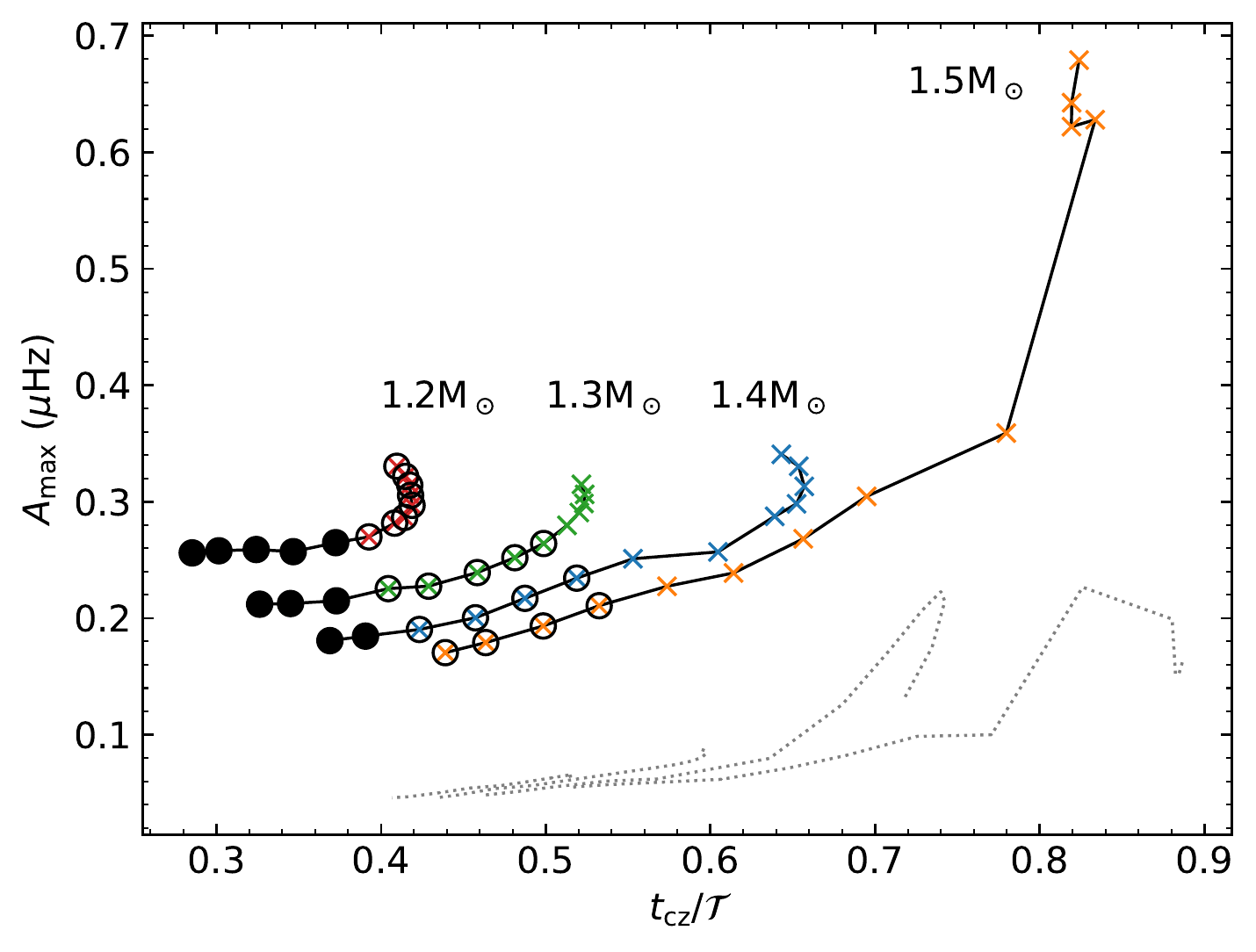}
    \includegraphics[scale=0.6]{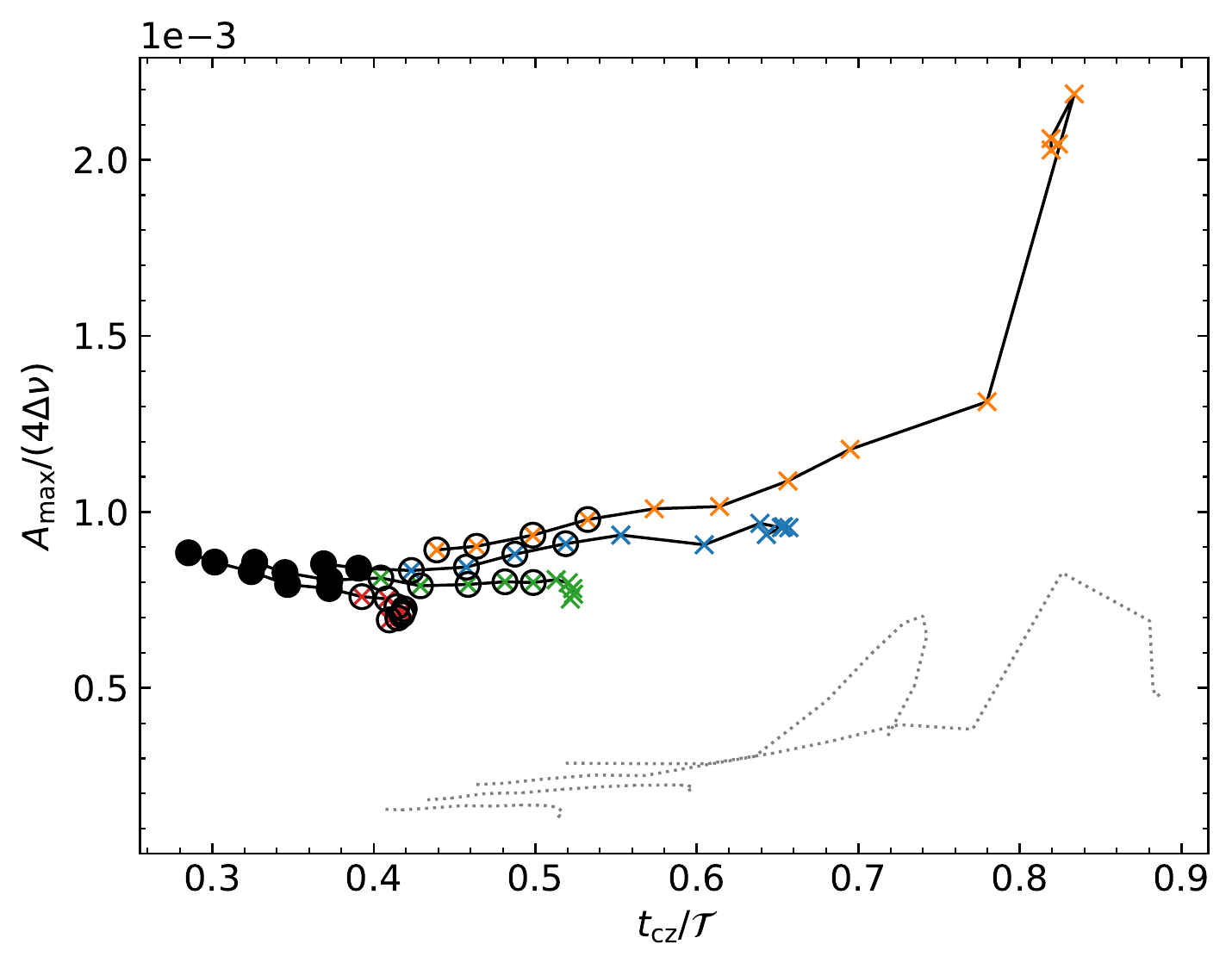}
    \includegraphics[scale=0.6]{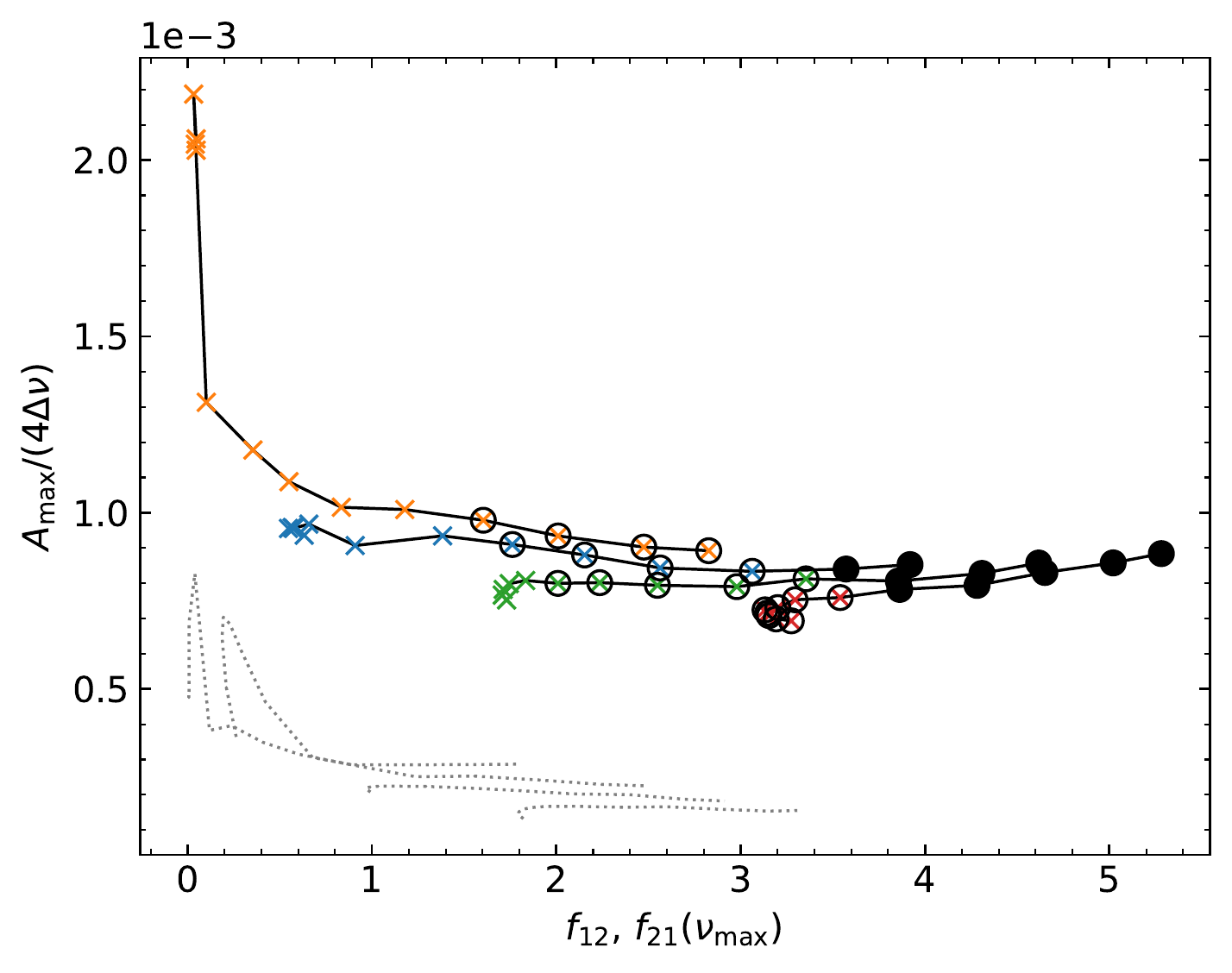}
   \caption{Detectability of the signature in the $r_{010}$ ratios for different quantities. The open and black circles show that the signature of the BSCZ glitch has an amplitude, $Amp=A_\mathrm{max} f_{12/21}(\nu_\mathrm{max})/(4\bar{\Delta})$, that is larger than $0.0015$ and $0.0030$ (the mean uncertainty of the $r_{010}$ ratios of 16 Cyg A and KIC10162436 around $\nu_\mathrm{max}$), respectively. \textbf{Top left:} Ratio of the acoustic radius of the BSCZ ($t_\mathrm{cz}$) over the total acoustic radius ($\mathcal{T}$) for models with masses between $1.2$ and $1.5$~M$_\odot$ and $\xi_{PC}=2$. The dashed line represents the position of the iron/nickel convective zone induced by their accumulations due to radiative accelerations for a 1.5~M$_\odot$ model. \textbf{Top right:} Associated evolutionary tracks in a Kiel diagram. \textbf{Middle left:} Amplitudes $A_\mathrm{max}$ according to the ratio of the acoustic radius of the BSCZ ($t_\mathrm{cz}$) over the total acoustic radius ($\mathcal{T}$) for the same models. \textbf{Middle right:} Amplitudes $A_\mathrm{max}/(4\Delta\nu)$ according to the ratio of the acoustic radius of the BSCZ ($t_\mathrm{cz}$) over the total acoustic radius ($\mathcal{T}$) for the same models. \textbf{Bottom:} Amplitudes $A_\mathrm{max}/(4\Delta\nu)$ according to $f_{12,21}$ at $\tilde{\nu}=\nu=\nu_\mathrm{max}$. For the top left, middle and bottom panels, the dotted grey lines represent the same models with $\xi_{PC}=0$.}
   \label{fig:7}
\end{figure*}
\subsection{Validation of the results}\label{valid:result}

For the validation, we considered the numerical frequencies of model B2 as our data and fit their $r_{010}$ ratios using Eq.~\ref{fit:2} and Eq.~\ref{fit:3}. In the fit, the functions $f_{12,21}/\bar \Delta$ were assumed to be known and were computed according to Eq.~\ref{eq:f12} and Eq.~\ref{eq:f21}. We now compare the amplitudes and acoustic radius resulting from the fit with the theoretical predictions computed according to Eqs.~\ref{eq:a1} and \ref{eq:a2}. An example of posterior distributions for the fitted parameters is shown in Appendix~\ref{appendix:B}.
The amplitudes $A_\mathrm{max}$ and the acoustic radius $t_d= t_\mathrm{cz}$ are well retrieved by the fitting procedure within the $1\sigma$ interval for most cases (and $2\sigma$ for all cases). Similarly to the results of \cite{monteiro94}, we find a degeneracy between $a_1$ and $a_2$ , as shown in Fig.~\ref{fig:C1}. The two fitting functions perform similarly for all cases, indicating that the function with the fewest parameters (Eq.~\ref{fit:3}) may be sufficient to analyse stars other than the Sun. Figure~\ref{fig:5} shows the predicted glitch signature from Eq.~\ref{eq:r010-th} for model B1 and B2.

For case 5, we considered a larger mean uncertainty of $0.30~\mu$Hz. The uncertainties are of the same order of magnitude as the amplitude, and the fitting of the glitch signature is more dependent on the prior imposed on $t_\mathrm{cz}$. In this case, we therefore adopted a two-step procedure. Firstly, we fit the signature without any informative prior. The distribution of $t_\mathrm{cz}$ is presented in Fig.~\ref{fig:6} for model B2 at $X_c=0.15$. Two peaks are clearly seen. One peak at large $t_\mathrm{cz}$ (close to the surface) is representative of a small-amplitude oscillation made possible by the large uncertainties and a second peak at the value extracted from the structure of the model. The strength of the latter depends on the ratio of the amplitude and the uncertainties. 
Secondly, we again performed the fit with informative priors ([3000; 4500]~s) around the significant peak. In this case, and when the second peak can be detected, the amplitude is retrieved much more clearly.

The reliability of the fit depends on the contrast between the amplitude of the signature and the uncertainties on the ratios. For $\xi_\mathrm{PC}<1.0$, the amplitude is often not large enough to be detected for uncertainties typical of F-type stars (e.g. $0.17~\mu$Hz for KIC10162436, which represents $0.0030$ for the ratios; see bottom panels of Fig.~\ref{fig:4}) with the input physics we considered in the models. An analysis of the detectability of the signature is deferred to Sect.~\ref{detect}. We first consider the case of the Sun as representative of a small PC region, but with highly precise frequencies in the following section.

\subsection{Sun as a test case}

We first tested our procedure on a solar-calibrated model with the same physics as model A0. The results are of the same quality level sd the cases presented in Sect.~\ref{valid:result}.

We next applied the one-step procedure to the BISON data of the Sun \citep{broomhall09,hale16}. We considered a prior on $t_\mathrm{cz}$ of [500; 2000]~s because the position of the BSCZ of the Sun is expected to be about $t_\mathrm{cz}/\mathcal{T}<0.5$ (with $\mathcal{T}=3701$~s obtained with $\Delta\nu_\mathrm{Sun}=135.1~\mu$Hz). The results of the fits are presented in Table~\ref{tab:4}. When compared to the results of \cite{monteiro94}, \cite{ballot04}, and \cite{JCD11}, the fitted value of $t_\mathrm{cz}$ lies between the values found in the three papers. The total amplitudes determined from the fits are similar for both fitting equations and similar to those found in previous studies. The differences found for the acoustic radius between the three determinations may come from the use of different solar frequency data sets (the \citealt{libbrecht90} data set was used in \citealt{monteiro94}, the six-year GOLF data set is used in \citealt{ballot04}, the \citealt{schou99} data set is used in \citealt{JCD11}, and the BISON data set is used in this work). Moreover, the value of the total acoustic radius of the Sun used to determine $t_\mathrm{cz}$ from $\tau_{cz}$ was obtained here using $\Delta\nu_\mathrm{Sun}=135.1~\mu$Hz, which may not be the same values as for the large separations used in the other studies. For example, a change of a few $\mu$Hz would reconcile the value of acoustic radii of the BSCZ. We checked that the two-step approach does not improve the agreement of $t_\mathrm{cz}$. The frequencies of the Sun are so precisely known that the approximations made in Appendix~\ref{appendix:A} maybe be too crude to analyse the solar signature of the glitch. The approach was also restricted to the use of $l=0$ and $1$ (because we used $r_{010}$ ratios) to mimic the conditions for stars other than the Sun, while the frequency variation method \citep[see][]{monteiro94} used as a comparison was applied with many more degrees ($l$ up to $20$). All this may explain the difference in $t_\mathrm{cz}$. We also stress that when compared to the results of \cite{roxburgh09}, which were also obtained with $r_{010}$ ratios, $t_\mathrm{cz}$ is very similar.

\section{Detectability of the signal in the $r_\mathrm{010}$ ratios}\label{detect}

Because the $r_{010}$ ratios are not expected to be sensitive to surface layers, the signature of the BSCZ is only detectable if the surface convective zone extends deep enough downward. The top left panel of Fig.~\ref{fig:7} shows the evolution of the ratio of the acoustic radius of the BSCZ ($t_\mathrm{cz}$) over the acoustic radius of the star ($\mathcal{T}$) for stellar models with masses between $1.2$ and $1.5$~M$_\odot$, with a solar initial chemical composition and a PC region of $\xi_{PC}=2.0$ (same input physics as model A5). In order to determine whether the signature of the glitch is detectable, we compared the total amplitude at $\nu=\nu_\mathrm{max}$ ($Amp=A_\mathrm{max} f_{12/21}(\nu_\mathrm{max})/(4\bar{\Delta})$) presented in Eq.~\ref{eq:r010_numax} with typical uncertainties of $0.0015$ and $0.0030$. The first uncertainty is the mean uncertainty of the ratios around $\nu_\mathrm{max}$ of 16 Cyg A, and the second uncertainty is the mean uncertainty of the ratios for frequencies between 0.85 and 1.15 $\nu_\mathrm{max}$ of KIC10162436. The detection threshold was set to $Amp/\sigma_\mathrm{ratio}=1$. This criterion is a convenient indication of detectability, it is sufficient but not necessary, as it does not mean that a detection cannot be made with a lower amplitude to the uncertainty ratios. Using this criterion, we remain on the conservative side. For the smallest ratio uncertainty, the signature of the BSCZ is detectable in the $r_{010}$ ratios according to the selected threshold if the BSCZ is deeper than $t_\mathrm{cz}/\mathcal{T}\approx0.55$. It extends below $t_\mathrm{cz}/\mathcal{T}\approx0.45$ for the largest uncertainty, which is representative of KIC10162436. This implies that in stars with a deep surface convective zone (M$\lesssim1.3$~M$_\odot$ at solar metallicity or for $X_c<0.3-0.2$ for more massive stars), the signature is more likely detectable. When the input physics is set, the detectability is directly linked to the effective temperature of the star ($T_\mathrm{eff}<6500$~K and $T_\mathrm{eff}<6183$~K for uncertainties of $0.0015$ and $0.0030$, respectively, for the models presented in Fig.~\ref{fig:7}). This can be explained by the fact that the size of the surface convective zone is directly correlated with $T_\mathrm{eff}$. The detection of a high amplitude glitch signature like this then provides strong constraints on the input physics of the models that control the lifetime of a star on the main sequence (e.g. overshoot or chemical transport close to the core).

Equation~\ref{eq:r010_numax} shows that the amplitude of the signal mainly depends on the product of $A_\mathrm{max}/(4\Delta\nu)$ and the functions $f_{12/21}$. The contributions to the amplitude of these terms are presented in the middle panels and in the bottom panel of Fig~\ref{fig:7}. For the signal to be detectable, the amplitude $A_\mathrm{max}$ needs to be larger than about 0.1~$\mu$Hz, but the contribution of $f_{12/21}$ also plays an important role.  $f_{12/21}$ are $\text{five}$ times larger at $t_\mathrm{cz}/\mathcal{T}=0.3$ than at $0.6$. In other words, the deeper the convection to radiation transition, the more likely the detection, almost independently of the value of $A_\mathrm{max}$, as shown in the bottom panel of Fig~\ref{fig:7}. This point also strengthens the constraining potential of this signal on the modelling of the stellar interior of these stars.

We also found that the changes in $\Gamma_1$ induced by the SAHA-S equation of state, or the sharp temperature gradient induced by an iron-nickel convective zone (see the dashed line in the top left panel of Fig.~\ref{fig:7}) appears too close to the surface to be detected. This explains the fact these changes do not lead to any signal in the $r_{010}$ ratios.

The top right panel of Fig~\ref{fig:7} shows the position of 3 \textit{Kepler} F-type stars in a Kiel diagram. The input physics of the models of the theoretical evolutionary tracks place KIC10162436 in the detectable region (considering uncertainties on the ratios of $0.003$), while the two other stars are placed in the detectable region considering uncertainties of the ratios of $0.0015$. These three stars will be analysed in a forthcoming paper. 

\section{Comparison with other seismic indicators}

\begin{table}
\centering
\caption{Results of the fit for cases 1, 3, and 6 of the frequency variation (freq.) and the second differences (diff.).} 
\label{sigs}
\begin{tabular}{ccc c cc}
\hline\hline
 $X_c$ & $\tilde{\nu}$ [$\mu$Hz] & $\bar{\Delta}$ [$\mu$Hz]  & Fit & $A_\mathrm{max}$ [$\mu$Hz] & $t_d$\tablefoottext{$\ast$} [s]  \\
\hline
%
 $0.25$  & $1232.6$ & $65.1$  & Th. & $0.232$ & $3931$  \\
\cmidrule(lr){4-6}
\smallskip
  &  &  & Freq. & $0.241^{+0.015}_{-0.016}$ & $3943^{+14}_{-17}$  \\
  &  &  & Diff. & $0.238^{+0.015}_{-0.015}$\tablefoottext{a} & $3935^{+16}_{-18}$  \\
\hline
%
 $0.15$  & $1091.4$ & $59.8$ & Th. & $0.204$ & $3757$  \\
\cmidrule(lr){4-6}
\smallskip
  &  &  & Freq. & $0.205^{+0.016}_{-0.016}$ & $3626^{+17}_{-16}$   \\  
  &  &  & Diff. & $0.206^{+0.016}_{-0.016}$\tablefoottext{a} & $3735^{+16}_{-20}$   \\
\hline
 $0.05$  & $982.2$ & $55.5$ & Th. & $0.190$ & $3575$    \\
\cmidrule(lr){4-6}
\smallskip
  &  &  & Freq. & $0.373^{+0.018}_{-0.018}$ & $4109^{+164}_{-916}$  \\
  &  &  & Diff. & $0.257^{+0.017}_{-0.018}$\tablefoottext{a} & $3438^{+26}_{-28}$  \\
\hline
\end{tabular}
\tablefoot{\tablefoottext{$\ast$}{Because both methods provide the acoustic depth $\tau_d$ of the glitch, the acoustic radius $t_d$ is calculated from $t_d=t_0-\tau_d$, with $t_0=1/(2\Delta\nu)$.}\tablefoottext{a}{The amplitudes are obtained by dividing $A_\mathrm{cz}^\ast$ by $ff_{21}(\nu_\mathrm{max}$ obtained for the second differences (see Appendix~\ref{appendix:C}).}}
\end{table} 

In this section, we analyse the same models as in the previous section (cases 1, 3, and 6) using other seismic indicators, namely the frequency variation \citep{monteiro94} and the second differences \citep[e.g.][]{gough90,monteiro93,verma19}. The fit of the signal was performed with the code called seismic inferences for glitches in stars (SIGS) presented in \cite{pereira17}.

The version of the code\footnote{SIGS for frequencies: \url{https://github.com/Fill4/sigs_freq}} for fitting the frequencies implements an automatic procedure for extracting the signal in $\nu_{n,\ell}$ through an iterative approach for removing the smooth component for each set of frequencies of degree $\ell$ and as a function of mode order $n$.
This implementation uses the method described by \cite{monteiro94} in Appendix~C, with the simplification done for low-degree data as used by \cite{monteiro00}.
The smooth component of the frequencies of the oscillations (what these would be in a star without a glitch) was iterated until it converged, and the final residuals were used to fit with the predicted expression for the signal.
\cite{pereira17} added an automatic and fairly robust approach to accomodate the need to have an initial guess and when the data are sparse and contain significant observational uncertainties.

The version of the code\footnote{SIGS for second differences: \url{https://github.com/Fill4/sigs_diff}} for fitting the second differences uses the approach from \cite{monteiro93}.
For low-degree data, a simplification was introduced by replacing the spline fits with the use of Eq.~\ref{eq:sec_dif}, which is similar to what was reported by \cite{gough90} and also used by several other authors (e.g.~\citealt{mazumdar14}, and references therein).
The approach removes a smooth component once from the second differences of the frequencies, defined as
\begin{equation}~\label{eq:sec_dif}
\Delta_2\nu_{n,\ell} \equiv \nu_{n{+}1,\ell} - 2 \nu_{n,\ell} + \nu_{n{-}1,\ell} \, ,
\end{equation}
and then fits the predicted expression of the signal to the residuals.

The details for both implementations and for the building of the automatic procedure are given in \cite{pereira17}. The method reported by \cite{pereira17} can use any variation in the expression for the signature in the frequencies and second differences. In their application, the lowest-amplitude case was considered by assuming no PC ($a_1$ is the only term used). This choice does not have a significant impact on the determination of the reference amplitude $A_{max}$, calculated at $\nu_{max}$. Because we here test the effect of a PC, the frequency dependence of the amplitude of the signature is better represented, as shown by \cite{monteiro94}, by including the $a_2$ term because $a_2 >> a_1$ in the presence of PC. The fitting functions (see Appendix~\ref{appendix:C} for the details on the signal for the second differences) are then
\begin{equation}\label{eq:signal_freqs}
  \begin{aligned}
      \delta\nu =& a_2 \left(\frac{\tilde{\nu}}{\nu}\right) \cos(4\pi\nu \tau_\mathrm{cz}+2\phi)\\
      &+ A_\mathrm{HeII}\left(\frac{\tilde{\nu}}{\nu}\right)\sin^2(2\pi\beta_\mathrm{HeII}\nu)\cos(4\pi\nu \tau_\mathrm{HeII}+2\phi_\mathrm{HeII}),\\
      \delta\Delta_2\nu =& A_\mathrm{cz}^\ast\left(\frac{\tilde{\nu}}{\nu}\right) \sin(4\pi\nu \tau_\mathrm{cz}+2\phi)\\
      &+ A_\mathrm{HeII}^\ast\left(\frac{\nu}{\tilde{\nu}}\right)\exp\left[-\beta_\mathrm{HeII}\left(\frac{\tilde{\nu}}{\nu}\right)^2\right]\sin(4\pi\nu \tau_\mathrm{HeII}+2\phi_\mathrm{HeII}).
  \end{aligned}    
\end{equation}
Both procedures in SIGS include the additional component, with an amplitude $A_{\rm HeII}$ that takes the glitch due to the second He ionisation into account.

In order to make the comparison of amplitudes meaningful, the amplitude of $\delta\Delta_2\nu$ associated with the BSCZ ($A_\mathrm{cz}^\ast$) is expressed according to $a_2$ following the same approach as in Appendix~\ref{appendix:A}. It can then be written
\begin{equation}\label{eq:sd}
\begin{aligned}
    \delta\Delta_2\nu_\mathrm{cz}&\approx a_2(\tau)\left(\frac{\tilde{\nu}}{\nu}\right)\left[\frac{2\nu^2}{\nu^2-\bar{\Delta}^2}\cos\left(4\pi\bar{\Delta}\tau_d\right)- 2\right]\cos\left(4\pi\nu\tau_\mathrm{cz}+2\phi\right)\\
    &\approx a_2(\tau)\left(\frac{\tilde{\nu}}{\nu}\right)\left[2-\frac{2\nu^2}{\nu^2-\bar{\Delta}^2}\cos\left(4\pi\bar{\Delta}\tau_d\right)\right]\sin\left(4\pi\nu\tau_\mathrm{cz}+2\phi'\right),
\end{aligned}
\end{equation}
\noindent where $\phi'=\phi-\pi/4$. $A_\mathrm{max}\approx a_2$ in the considered cases because the models include PC. $A_\mathrm{max}$ can then be retrieved from $A_\mathrm{cz}^\ast$ using the amplitude of Eq.~\ref{eq:sd}.

The results of the fit of the signals for the three cases are presented in Table~\ref{sigs}. For cases 1 and 3, the position of the glitch is well retrieved by both seismic indicators. The amplitude is also well recovered with the fit of the frequency variation. Nevertheless, the deeper the glitch (i.e. the more evolved the star), the less efficiently retrieve these two indicators the glitch properties because the amplitude of the signature is smaller (see Fig.~\ref{fig:7} for the variation of $A_\mathrm{max}$ with depth), especially for the frequency fit. For the second differences, the amplification of the amplitude by the frequency dependence is maximum for BSCZ at $t_\mathrm{cz}/\mathcal{T}=0.5$ and decreases when the BSCZ deepens (see Fig.~\ref{fig:C1}), which is consistent with the results of \cite{ballot04}. This explains why the position and amplitude of the glitch signature for case 6 are not well retrieved by the second differences either.

Accordingly, for G-type stars with highly precise frequencies (e.g. the Sun), the three methods ought to perform equivalently to retrieve the properties of the glitch signature induced by the BSCZ. For stars with   lower frequency precision, the second differences and the ratios $r_{010}$ are expected to perform better than the frequency variation. The efficiency of these last two methods depends on the position of the glitch. Deeper glitches ($t_\mathrm{cz}/\mathcal{T}<0.4$) should be better retrieved by the ratios, and the second differences should perform better for $t_\mathrm{cz}/\mathcal{T}>0.4$. For F-type stars, the conclusions are the same.

\section{Impact of a smoother transition and a $\mu$ gradient}

\begin{figure*}[t]
    \centering
    \includegraphics[scale=0.6]{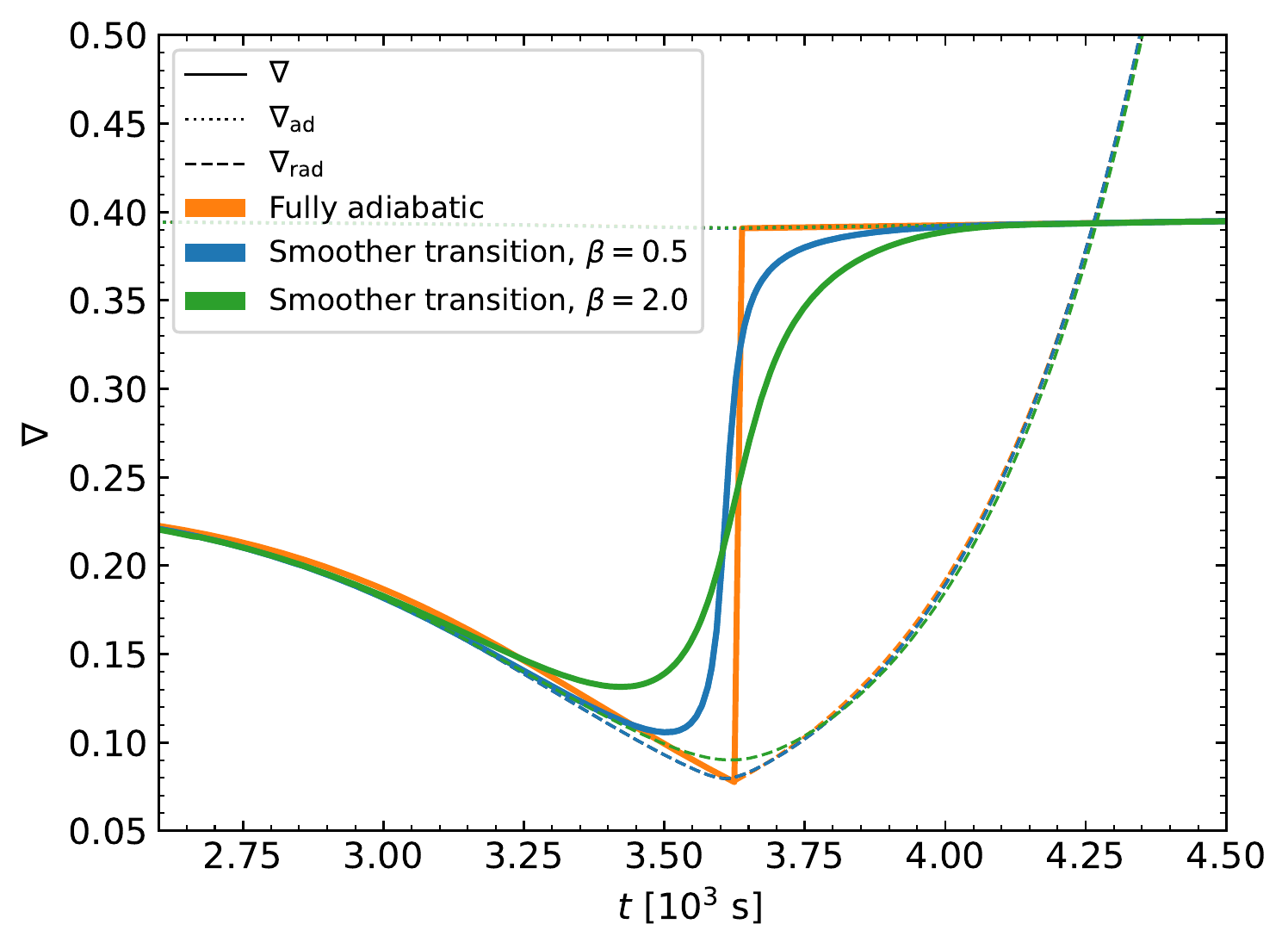}
    \includegraphics[scale=0.6]{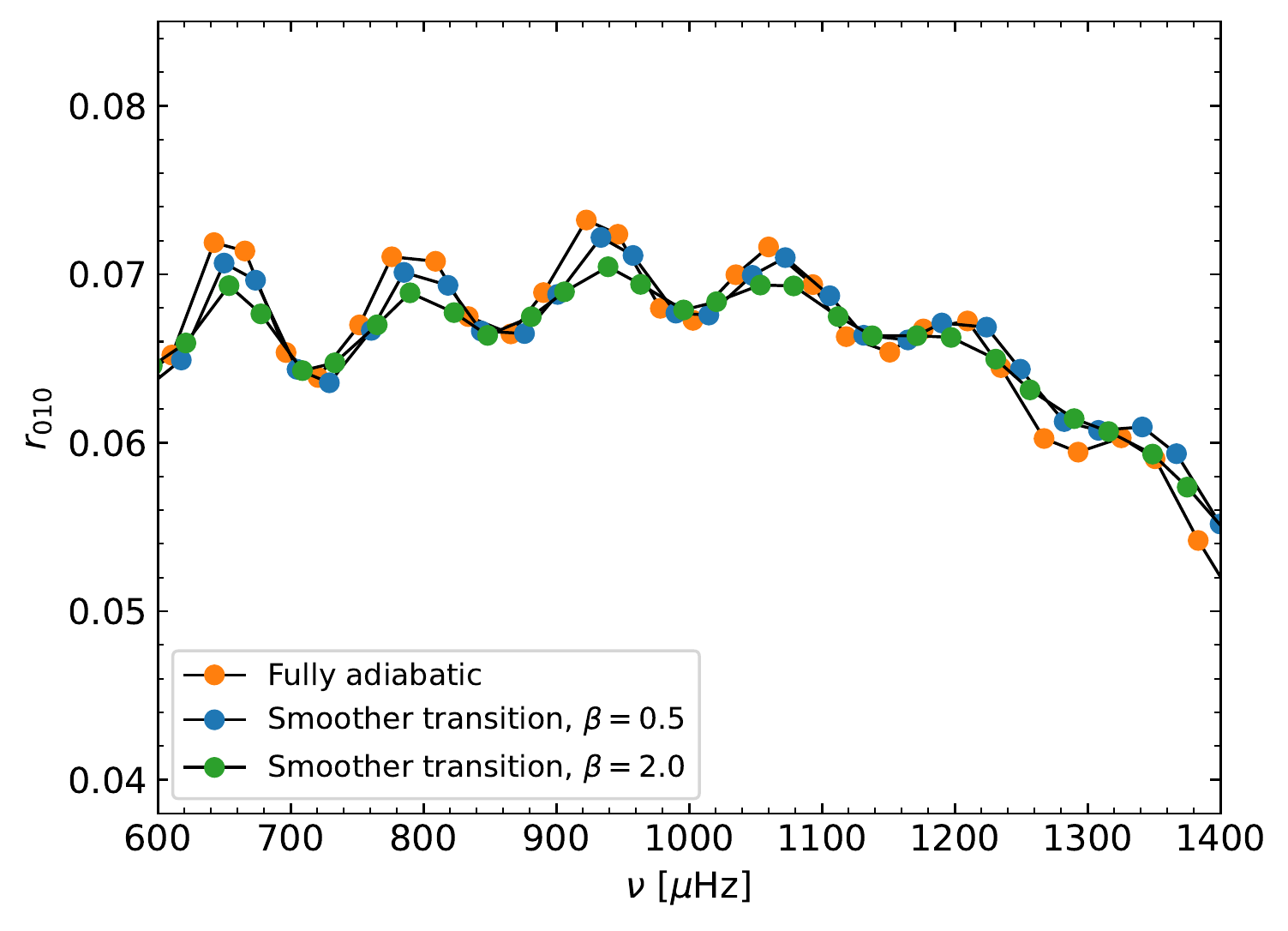}
    \includegraphics[scale=0.6]{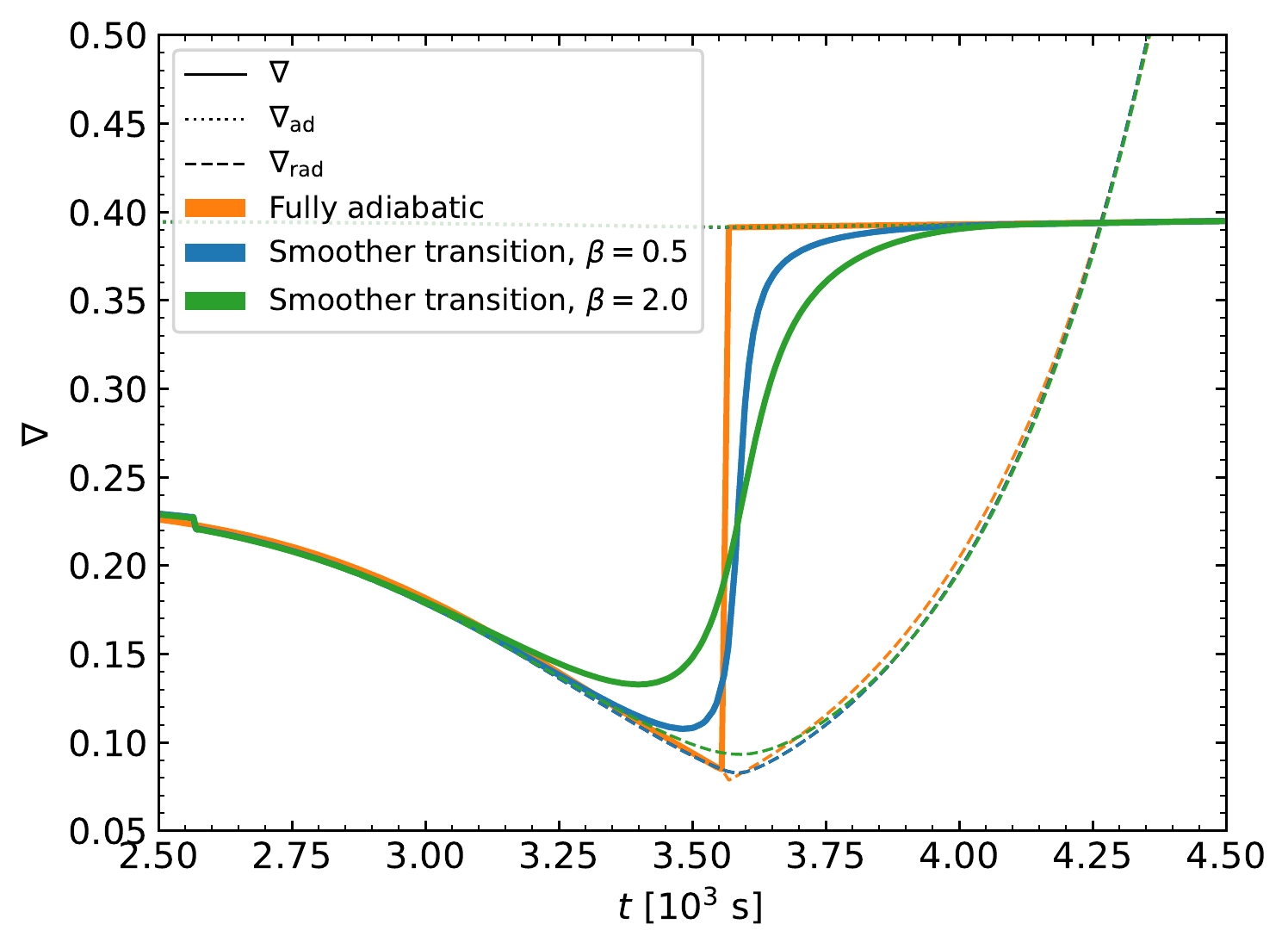}
    \includegraphics[scale=0.6]{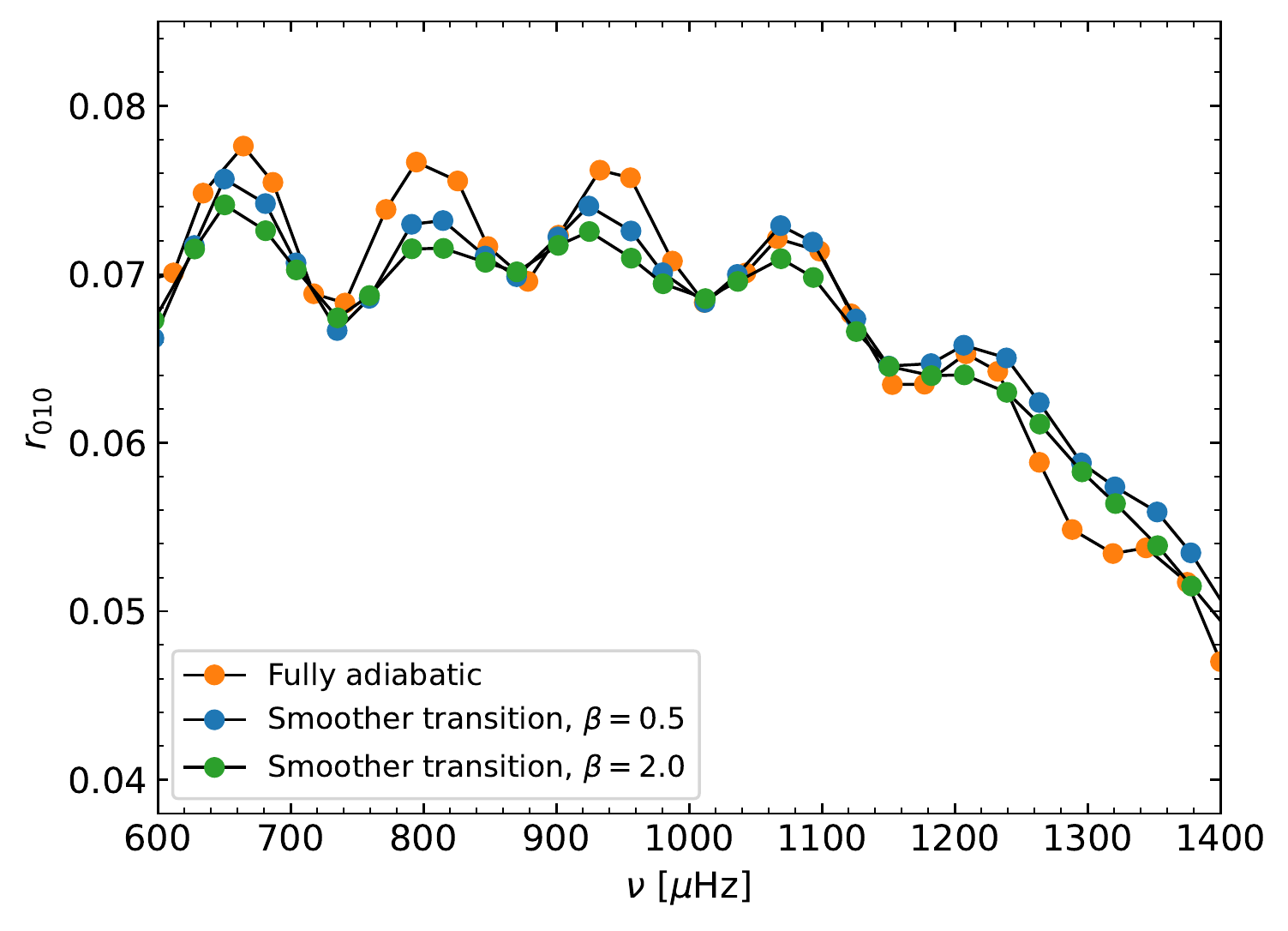}
   \caption{Impact of the $\mu$-gradient and of the smoothness of the temperature gradient on the $r_{010}$ ratios. \textbf{Left}: Temperature gradients of model B2 (orange), B3 (blue), and B4 (green) for the top panel, and A5 (orange), A6 (blue), and A7 (green) for the bottom panel as a function of the acoustic radius at $X_c=0.10$. \textbf{Right}: Corresponding $r_{010}$ ratios as a function of the frequency.}
   \label{fig:2bis}
\end{figure*}

The top panels of Fig.~\ref{fig:2bis} show the comparisons of the temperature gradients and $r_{010}$ ratios for models B2, B3, and B4 at $X_c=0.10$. The smoother transition of models B3 and B4 occurs at the same location as the bottom of the PC region of model B2. The fitted amplitudes (with a mean uncertainty on the frequency of $0.10~\mu$Hz and using Eq.~\ref{fit:2}) are $0.172^{+0.023}_{-0.022}$, $0.140^{+0.021}_{-0.021}$ and $0.080^{+0.021}_{-0.019}~\mu$Hz for models B2, B3, and B4, respectively. The distinction between the two types of temperature gradients may be possible with precise frequencies ($\sigma_\mathrm{freq}\approx0.10~\mu$Hz), but would be more difficult for larger uncertainties (e.g. typical error for F-type stars).

When atomic diffusion is taken into account, a $\mu$ gradient is built at the bottom of the PC region. We assumed here that the mixing in the PC region was the same as in the convective zone down to the position at which the temperature gradient is radiative (below the transition, around $2600$~s). The bottom panels of Fig.~\ref{fig:2bis} show the comparisons of the temperature gradients and $r_{010}$ ratios for models A5, A6, and A7 at $X_c=0.10$. The smoother transition of models A6 and A7 occurs at the same location as the bottom of the PC region of model A5. The fitted amplitudes (with a mean uncertainty on the frequency of $0.10~\mu$Hz) are $0.177^{+0.026}_{-0.026}$, $0.131^{+0.021}_{-0.020}$ and $0.077^{+0.027}_{-0.024}~\mu$Hz for models A5, A6, and A7, respectively.

For model A5, the amplitude appears to be amplified (see the discussion in Sect.~\ref{sect4.1} about the impact of a $\mu$ gradient). We note that in addition to the BSCZ glitch signature, a second signature seems to distort the signal for the models with atomic diffusion (i.e. models A6 and A7 with a $\mu$ gradient at a different position than the BSCZ). This might indicate that the signal is composed of the signature of the BSCZ and a deeper $\mu$ gradient. This needs confirmation, however.

\section{Discussion and conclusion}\label{discu-conclu}
 
We have theoretically studied the diagnostic power of the ratios $r_{010}$ to probe the depth of the surface convective zone of F-type stars. The oscillations of these ratios indeed present  very large amplitudes  that are not observed in G-type stars, that is, stars with  lower mass, and especially, cooler stars with deeper convective envelopes.

Our numerical and analytical investigations confirm that convective penetration below the convective envelope is able to produce the large-amplitude variations of the glitch signature that are observed in the frequency ratios of solar-like F-type stars. Stellar models including a large extension of the convective penetration indeed produce large-amplitude BSCZ glitch signatures as observed and the $r_{010}$ ratios. 
These signatures provide the location for the base of the extended convective region in the models. No other physical effect investigated here is able to explain these observations.
For the F-type star models and the input physics we considered in this study, the large amplitudes can indeed only be achieved when the extent is sufficiently deep ($1-2~H_p$) so as not to be filtered by the ratios of frequencies. The frequency dependence of the amplitude also amplifies the detectability of the signature according to the depth of the transition between the convective envelope and the radiative zone below. However, the depth of the PC region exceeds that measured for the Sun. For models with other input physics (e.g. initial chemical composition or transport processes), the extent of the PC region needed to explain the signature may be different. Hence, this high value may be an indication that the input physics of the models is still incomplete, which leads to an underestimation of the size of convective envelopes, or it might indicate that another mechanism causes the signature. This should be assessed by modelling of the F-type stars of the \textit{Kepler} Legacy sample, with additional physics that may affect the BSCZ.

Although convective penetration is also expected in lower-mass, cooler stars such as the Sun, no such large amplitudes are observed. Assuming then that PC causes the large-amplitude signature of the $r_{010}$ ratios, we find that the transition between the convective envelope to the radiative interior is smoother and shallower for less massive stars. For the Sun, we also found that the free extension parameter $\xi$ in the Zahn (1991) formalism is no longer unity when updated microphysics is used, but is rather a factor of ten lower. This tends to indicate that this parameter depends on the physics of the models and that the Zahn (1991) formalism needs to be developed further.

The analysis of the signature of the BSCZ in ratios $r_{010}$ was performed (similarly to previous works) without the contribution of the $\mu$ gradient. We nevertheless found that it may impact the oscillatory signature of the ratios, hence the characterisation of the glitch properties. The second paper of this series will indeed be devoted to investigate the contribution of the $\mu$ gradient for a deeper understanding of the glitch signatures seen in the ratios $r_{010}$.

Finally, our study emphasised that the ratios of small to large separations can be used for stars other than the Sun to determine valuable constraints on stellar interiors, despite larger uncertainties on the frequencies. We also showed that in some specific cases, the ratios $r_{010}$ allow us to better recover the position of the BSCZ than the second differences or the frequencies themselves. Much can be done with the $l=0$ and $1$ modes alone, which can be easily accessible for a large number of solar-like oscillating main-sequence stars. This opens new interesting possibilities to test 3D simulation predictions of temperature gradients in the penetration convection regions outside of the solar parameter space \citep[e.g.][]{anders22, breton22} and to place constraints on the physical conditions of these regions.
 
\begin{acknowledgements}
We gratefully thank our anonymous referee whose comprehensive readings and remarks helped to improve the content of the manuscript. We also thank Jérôme Ballot for an interesting discussion related to this work. This work was supported by CNES, focused on PLATO. This work was supported by FCT/MCTES through the research grants UIDB/04434/2020, UIDP/04434/2020 and PTDC/FIS-AST/30389/2017. MD and MSC (CEECIND/02619/2017) are supported by national funds through FCT in the form of a work contract. GB acknowledges fundings from the SNF AMBIZIONE grant No 185805 (Seismic inversions and modelling of transport processes in stars). MD thanks Gaëtan Sary for fruitful mathematical discussions.
\end{acknowledgements}  

\bibliographystyle{aa} 
\bibliography{main.bib} 
\newpage
\begin{appendix}
\section{Expression of the $r_{010}$ ratios}\label{appendix:A}

\begin{figure}[t]
    \centering
    \includegraphics[scale=0.6]{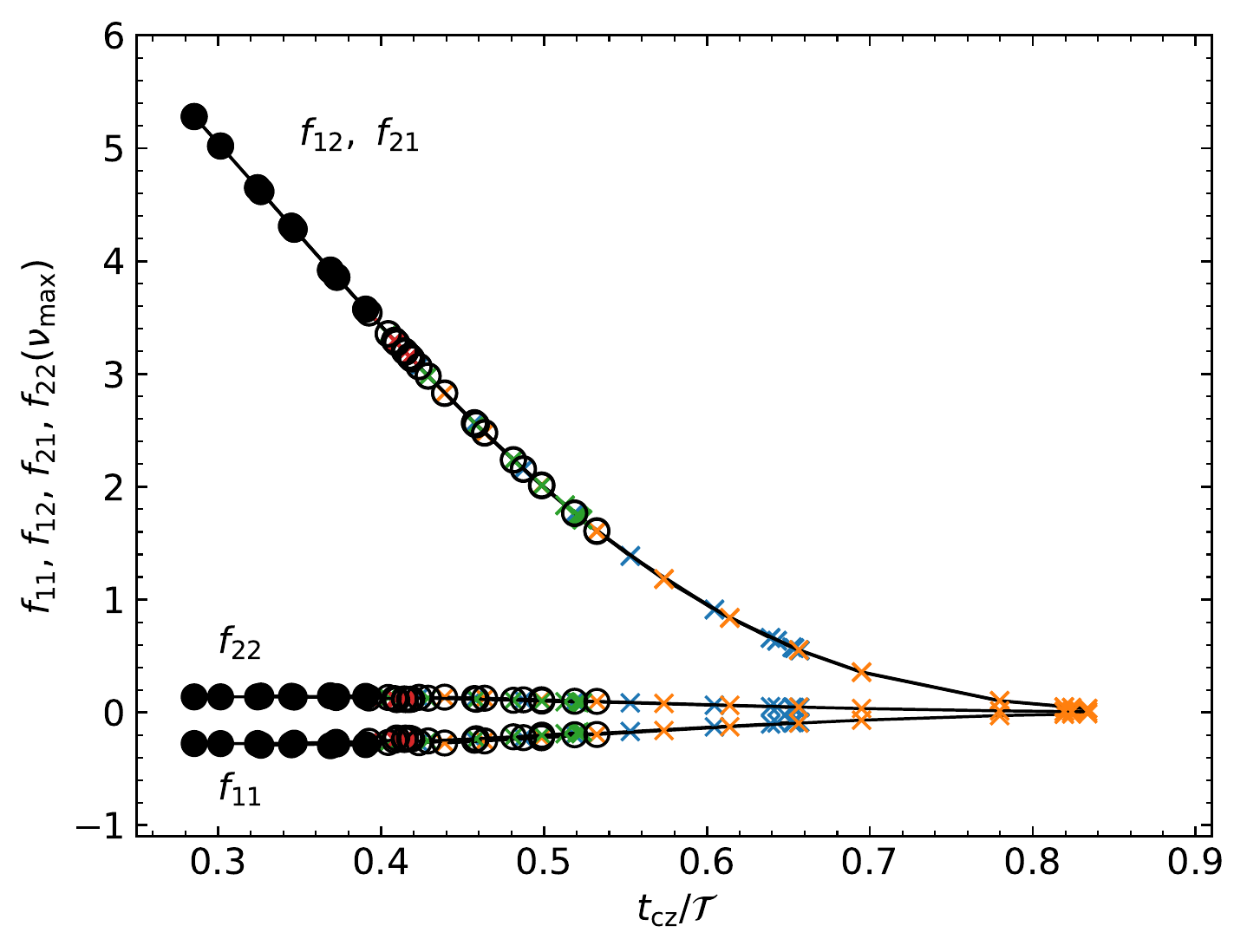}
    \caption{Value of the $f$ functions at $\tilde{\nu}=\nu=\nu_\mathrm{max}$  for the same  models as in Fig.~\ref{fig:7}, i.e. with masses between $1.2$ and $1.5$~M$_\odot$ and $\xi_{PC}=2$.  }
   \label{fig:A1}
\end{figure}

We first write all the frequencies needed to calculate $d_{01}$ and $d_{10}$ with a constant frequency ($\tilde{\nu_{n,l}}$) plus a small variation,
\begin{align}
    \nu_{n,l}=\tilde{\nu_{n,l}}+\delta\nu_{n,l}~,
\end{align}
and assuming a mean large separation over $\bar{\Delta}$ for angular degree $0$ and $1,$ we obtain
\begin{align}
    \nu_{n+1,l}=&\nu_{n,l}+\bar{\Delta}~,\\
    \nu_{n-1,l}=&\nu_{n,l}-\bar{\Delta}~.
\end{align}
The frequencies needed to calculate the small separations are then defined for $l=0,1$ by 
\begin{equation}
  \begin{aligned}
    \nu_{n,l} &=\tilde{\nu_{n,l}}+\delta\nu(\nu_{n,l}),\\
    \nu_{n+1,l}&=\tilde{\nu_{n,l}}+\bar{\Delta}+\delta\nu(\nu_{n,l}+\bar{\Delta}),\\
    \nu_{n-1,l}&=\tilde{\nu_{n,l}}-\bar{\Delta}+\delta\nu(\nu_{n,l}-\bar{\Delta}).\\
  \end{aligned}
\end{equation}
Using the asymptotic expression of the frequency
\begin{align}
    \nu_{n,l}\approx\left(n+\frac{l}{2}+\frac{1}{4}+\alpha\right)\bar{\Delta},
\end{align}
\noindent we can define the relation between $l=0$ and $l=1$ modes by
\begin{align}
    \nu_{n,1}\approx\nu_{n,0}+\frac{\bar{\Delta}}{2}.
\end{align}
We can then express the five frequencies required to estimate the $r_{01}$ ratios according to the frequency $\nu_{n,0,}$
\begin{equation}
  \begin{aligned}
    \nu_{n,0}&=\tilde{\nu_{n,0}}+\delta\nu(\nu_{n,0}),\\
    \nu_{n+1,0}&=\tilde{\nu_{n,0}}+\bar{\Delta}+\delta\nu(\nu_{n,0}+\bar{\Delta}),\\
    \nu_{n-1,0}&=\tilde{\nu_{n,0}}-\bar{\Delta}+\delta\nu(\nu_{n,0}-\bar{\Delta}),\\
    \nu_{n,1}&=\tilde{\nu_{n,0}}+\frac{\bar{\Delta}}{2}+\delta\nu(\nu_{n,0}+\bar{\Delta}/2),\\
    \nu_{n-1,1}&=\tilde{\nu_{n,0}}-\frac{\bar{\Delta}}{2}+\delta\nu(\nu_{n,0}-\bar{\Delta}/2).\\ 
  \end{aligned}
\end{equation}
When this is injected in Eq.~\ref{eq:r010}, we obtain
\begin{equation}\label{eq:A8}
    \begin{aligned}
        r_{01}(n)=&\frac{1}{8\bar{\Delta}}\left[\delta\nu(\nu_{n,0}-\bar{\Delta})-4\delta\nu(\nu_{n,0}-\frac{\bar{\Delta}}{2})+6\delta\nu(\nu_{n,0})\right. \\
       &\left.-4\delta\nu(\nu_{n,0}+\frac{\bar{\Delta}}{2})+\delta\nu(\nu_{n,0}+\bar{\Delta})\right].\\
    \end{aligned}
\end{equation}

We now assume that the small frequency departure is due to  the BSCZ glitch and use Eq.~\ref{eq:mario} for $\delta \nu$. In the following, we replace $\nu_{n,0}$ by $\nu$ for sake of clarity,
\begin{equation}\label{eq:A9}
    \begin{aligned}
        \delta\nu(\nu\pm\bar{\Delta})=&a_2(\tau_d)\left(\frac{\tilde{\nu}}{\nu\mp\bar{\Delta}}\right)\left[\cos\left(4\pi\nu\tau_d+2\phi\right)\cos\left(4\pi\bar{\Delta}\tau_d\right)\right.\\
        &\left.\mp\sin\left(4\pi\nu\tau_d+2\phi\right)\sin\left(4\pi\bar{\Delta}\tau_d\right)\right]\\
        &+a_1(\tau_d)\left(\frac{\tilde{\nu}}{\nu\pm\bar{\Delta}}\right)^2\left[\sin\left(4\pi\nu\tau_d+2\phi\right)\cos\left(4\pi\bar{\Delta}\tau_d\right)\right.\\
        &\left.\pm\cos\left(4\pi\nu\tau_d+2\phi\right)\sin\left(4\pi\bar{\Delta}\tau_d\right)\right],
    \end{aligned}
\end{equation}
\begin{equation}\label{eq:A10}
    \begin{aligned}
        \delta\nu(\nu\pm\frac{\bar{\Delta}}{2})=&a_2(\tau_d)\left(\frac{\tilde{\nu}}{\nu\pm\frac{\bar{\Delta}}{2}}\right)\left[\cos\left(4\pi\nu\tau_d+2\phi\right)\cos\left(2\pi\bar{\Delta}\tau_d\right)\right.\\
        &\left.\mp\sin\left(4\pi\nu\tau_d+2\phi\right)\sin\left(2\pi\bar{\Delta}\tau_d\right)\right]\\
        &+a_1(\tau_d)\left(\frac{\tilde{\nu}}{\nu\pm\frac{\bar{\Delta}}{2}}\right)^2\left[\sin\left(4\pi\nu\tau_d+2\phi\right)\cos\left(2\pi\bar{\Delta}\tau_d\right)\right.\\
        &\left.\pm\cos\left(4\pi\nu\tau_d+2\phi\right)\sin\left(2\pi\bar{\Delta}\tau_d\right)\right].
    \end{aligned}
\end{equation}

When we insert Eq.~\ref{eq:A9} and \ref{eq:A10} into Eq.~\ref{eq:A8}, 
the $r_{01}$ ratios then become
\begin{equation}
    \begin{aligned}
        r_{01}(\nu)=&\frac{a_2(\tau_d)}{4\bar{\Delta}}\left[\left(\frac{\tilde{\nu}\nu}{\nu^2-\bar{\Delta}^2}\right)\cos\left(4\pi\bar{\Delta}\tau_d\right)-\left(\frac{4\tilde{\nu}\nu}{\nu^2-\bar{\Delta}^2/4}\right)\cos\left(2\pi\bar{\Delta}\tau_d\right)\right. \\
       &\left.+3\left(\frac{\tilde{\nu}}{\nu}\right)\right]\times\cos\left(4\pi\nu\tau_d+2\phi\right)\\
       +&\frac{a_2(\tau_d)}{4\bar{\Delta}}\left[\left(\frac{\tilde{\nu}\bar{\Delta}}{\nu^2-\bar{\Delta}^2}\right)\sin\left(4\pi\bar{\Delta}\tau_d\right)-\left(\frac{2\tilde{\nu}\bar{\Delta}}{\nu^2-\bar{\Delta}^2/4}\right)\sin\left(2\pi\bar{\Delta}\tau_d\right)\right] \\
       &\times\sin\left(4\pi\nu\tau_d+2\phi\right)\\
       +&\frac{a_1(\tau_d)}{2\bar{\Delta}}\left[-\left(\frac{\tilde{\nu}^2\nu\bar{\Delta}}{(\nu^2-\bar{\Delta}^2)^2}\right)\sin\left(4\pi\bar{\Delta}\tau_d\right)\right.\\
       &\left.+\left(\frac{2\tilde{\nu}^2\nu\bar{\Delta}}{(\nu^2-\bar{\Delta}^2/4)^2}\right)\sin\left(2\pi\bar{\Delta}\tau_d\right)\right]\times\cos\left(4\pi\nu\tau_d+2\phi\right)\\
       +&\frac{a_1(\tau_d)}{4\bar{\Delta}}\left[\left(\frac{\tilde{\nu}^2(\nu^2+\bar{\Delta}^2)}{(\nu^2-\bar{\Delta}^2)^2}\right)\cos\left(4\pi\bar{\Delta}\tau_d\right)\right.\\
       &\left.-\left(\frac{4\tilde{\nu}^2(\nu^2+\bar{\Delta}^2/4)}{(\nu^2-\bar{\Delta}^2/4)^2}\right)\cos\left(2\pi\bar{\Delta}\tau_d\right)+3\left(\frac{\tilde{\nu}}{\nu}\right)^2\right]\\
       &\times\sin\left(4\pi\nu\tau_d+2\phi\right).
    \end{aligned}
\end{equation}
Similarly, centring the expression around the $\nu_{n,1}$ and replacing it by $\nu$ for sake of clarity, we obtain the same expression for $r_{01}(\nu)$. In the following, we then refer to $r_{010}(\nu)$ as the combination of $r_{10}(\nu)$ and $r_{01}(\nu)$.

Because the period of the signature in the $r_{010}$ ratios is the acoustic radius rather than the acoustic depth \citep{roxburgh09}, we convert the previous expression in terms of $t_d=\mathcal{T}-\tau_d$ instead of $\tau_d$ (using $\mathcal{T}=1/(2\bar{\Delta})$). The $r_{010}$ ratios are then given by
\begin{equation}\label{eq:r010_A}
    \begin{aligned}
        r_{010}(\nu)=&\frac{a_1(\tau_d)}{4\bar{\Delta}}\left(\frac{\tilde{\nu}}{\nu}\right)^2\left[f_{11}(\nu)\times\cos\left(4\pi\nu t_d+2\phi\right)+\right.\\
        &\left.+f_{12}(\nu)\times\sin\left(4\pi\nu t_d+2\phi\right)\right]\\
        +&\frac{a_2(\tau_d)}{4\bar{\Delta}}\left(\frac{\tilde{\nu}}{\nu}\right)\left[ f_{21}(\nu)\times\cos\left(4\pi\nu t_d+2\phi\right)\right.\\
        &\left.+f_{22}(\nu)\times\sin\left(4\pi\nu t_d+2\phi\right)\right],
    \end{aligned}
\end{equation}
with
\begin{equation}
    \begin{aligned}
        f_{11}(\nu)=-\left(\frac{2\nu^3\bar{\Delta}}{(\nu^2-\bar{\Delta}^2)^2}\right)\sin\left(4\pi\bar{\Delta}\tau_d\right)-\left(\frac{4\nu^3\bar{\Delta}}{(\nu^2-\bar{\Delta}^2/4)^2}\right)\sin\left(2\pi\bar{\Delta}\tau_d\right),
    \end{aligned}
\end{equation}
\begin{equation}
    \begin{aligned}
        f_{12}(\nu)=&\left[\frac{\nu^2(\nu^2+\bar{\Delta}^2)}{(\nu^2-\bar{\Delta}^2)^2}\right]\cos\left(4\pi\bar{\Delta}t_d\right)\\
        +& \left[\frac{4\nu^2(\nu^2+\bar{\Delta}^2/4)}{(\nu^2-\bar{\Delta}^2/4)^2}\right]\cos\left(2\pi\bar{\Delta}t_d\right)
        +3,
    \end{aligned}
\end{equation}
\begin{equation}
    \begin{aligned}
        f_{21}(\nu)=&\left[\frac{\nu^2}{\nu^2 - \bar{\Delta}^2}\right]\cos\left(4\pi\bar{\Delta}t_d\right)\\
        +& \left[\frac{4\nu^2}{\nu^2-\bar{\Delta}^2/4}\right]\cos\left(2\pi\bar{\Delta}t_d\right)+3,
    \end{aligned}
\end{equation}
\begin{equation}
    \begin{aligned}
        f_{22}(\nu)=    \left(\frac{\nu\bar{\Delta}}{\nu^2-\bar{\Delta}^2}\right)\sin\left(4\pi\bar{\Delta}\tau_d\right)+\left(\frac{2\nu\bar {\Delta}}{\nu^2-\bar{\Delta}^2/4}\right)\sin\left(2\pi\bar{\Delta}\tau_d\right).
    \end{aligned}
\end{equation}
For the  excited solar-like oscillation modes, the ratio 
$$  {\cal R}\equiv \bar \Delta /\nu  \approx \frac{1}{n+l/2+\epsilon} <<1~. $$
This is confirmed in Table~\ref{tab:3}, which lists  ${\cal R} \sim 0.05 <<1, \text{which is}$ roughly  the same for the considered stellar models. Then at first order,
 \begin{equation}
  \begin{aligned}
     f_{11}(\nu)\approx-  2   {\cal R}   
        \Bigl(  {\cal  R}      ~  \sin\left(2 \Phi\right)
        + 2    ~ \sin\left(\Phi\right) \Bigr),
    \end{aligned}
  \end{equation}
\begin{equation}
    \begin{aligned}
        f_{12}(\nu) \approx    f_{21}(\nu) \approx &   ~3+
          ~ \cos\left(2\Phi_t\right)
        + 4   ~\cos\left(\Phi_t\right),
    \end{aligned}
\end{equation}
\begin{equation}
    \begin{aligned}
        f_{22}(\nu) &\approx  {\cal R}  \Biggl( \Biggr.
    ~\sin\left(2 \Phi\right)
        +2   ~ \sin\left(\Phi\right)  \Biggl. \Biggr),  \\  
    \end{aligned}
\end{equation}
\noindent where $\Phi_t= 2\pi\bar{\Delta} t_d $ and $\Phi = 2\pi\bar{\Delta}\tau_d $.\\

We consider the ratios
   \begin{equation} 
  \frac{f_{11}}{f_{12}}\approx  -  4   {\cal R} ~\frac{ \sin\left(\Phi\right)}{3+\cos\left(2\Phi_t\right)
        + 4   ~\cos\left(\Phi_t\right)}.
   \end{equation}

Table~\ref{tab:3} shows that $\Delta t_d \sim 1/4$. This means $\Phi_t= 2\pi\bar{\Delta}  t_d  \sim   2\pi/4 =\pi/2 $. On the other hand, $\Phi = 2\pi\bar{\Delta}\tau_d = 2\pi \bar{\Delta} (\mathcal{T}-t_d) = \pi -  2\pi \bar{\Delta}t_d  = \pi/2  \approx \Phi_t$ , then  

\begin{equation} 
  \frac{f_{11}}{f_{12}}\approx  -  4   {\cal R} ~
  \frac{ \sin\left(\Phi\right)}{3+ \cos\left(2\Phi_t\right)
        +  4  ~   \cos\left(\Phi_t\right)} 
  \approx -2 {\cal R} \approx -0.1~,
 \end{equation}
\begin{eqnarray}
\frac{f_{22}}{f_{21}} \approx  2 
{\cal R}  ~\frac{  \sin\left(\Phi\right) \Bigl(  \cos \left(\Phi\right) +1\Bigr)  }{ 3+ \cos\left(2\Phi_t\right)
        +  4  ~   \cos\left(\Phi_t\right)}  \approx   {\cal R} \approx 0.05~.
\end{eqnarray}
 
This indicates that $f_{11}, f_{22}$ are negligible compared to $f_{12},f_{21}$ , respectively. This is illustrated in Fig.~\ref{fig:A1}, which shows the $f_{ij}$ coefficients. 

\onecolumn

\section{Fitting of $a_1$, $a_2$, $A_\mathrm{max}$ , and $t_d$}\label{appendix:B}

\begin{figure*}[!ht]
    \centering
    \includegraphics[scale=0.6]{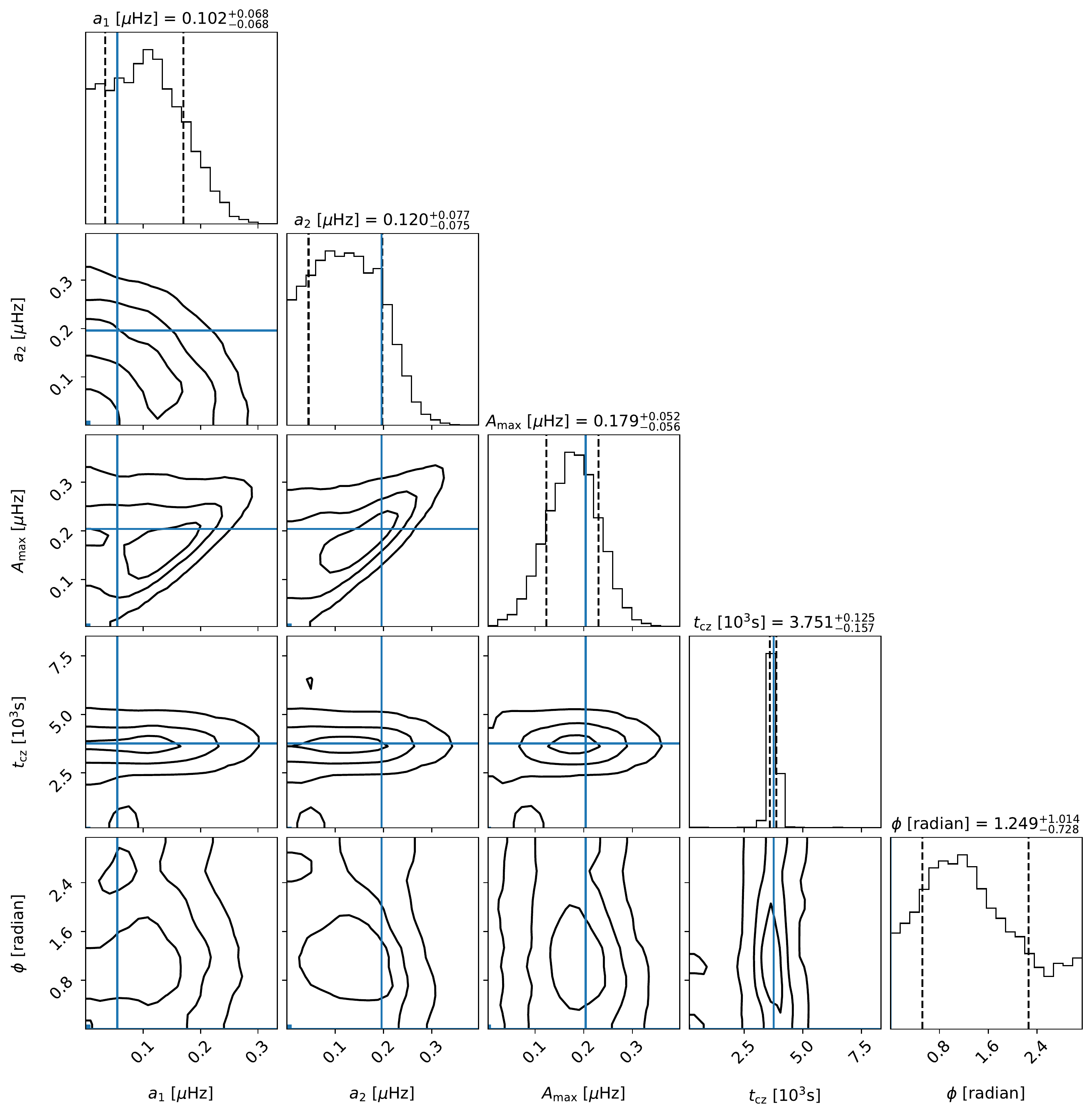}
   \caption{Case 4 with Eq.~\ref{fit:2} of Table~\ref{tab:3}.}
   \label{fig:B1}
\end{figure*}

Figure~\ref{fig:B1} shows the resulting parameters of the fit of the $r_{010}$ signature of model B2 using Eq.~\ref{fit:2}.

\twocolumn
\newpage
\section{Expression of $\Delta_2\nu$ according to $a_1$ and $a_2$}\label{appendix:C}

The second differences are defined by
\begin{equation}
    \Delta_2\nu_{n,l}= \nu_{n-1,l}-2\nu_{n,l}+\nu_{n+1,l}
.\end{equation}

Following the same approach as in Appendix~\ref{appendix:A}, the signature of the BSCZ in the second differences is expressed as 
\begin{equation}
    \begin{aligned}
          \Delta_2\nu_{n,l,cz}=&a_1(\tau)\left(\frac{\tilde{\nu}}{\nu}\right)^2[ff_{11}(\nu)\cos\left(4\pi\nu\tau_d+2\phi\right)\\
          &+ff_{12}(\nu)\sin\left(4\pi\nu\tau_d+2\phi\right)]\\
          &+a_2(\tau)\left(\frac{\tilde{\nu}}{\nu}\right)[ff_{21}(\nu)\cos\left(4\pi\nu\tau_d+2\phi\right)\\
          &+ff_{22}(\nu)\sin\left(4\pi\nu\tau_d+2\phi\right)]\\
    \end{aligned}
\end{equation}
\noindent with
\begin{equation}
    ff_{11}(\nu)=-\frac{4\nu^3\bar{\Delta}}{(\nu^2-\bar{\Delta}^2)^2}\sin\left(4\pi\bar{\Delta}\tau_d\right)
\end{equation}
\begin{equation}
    ff_{12}(\nu)=\frac{2\nu^2(\nu^2+\bar{\Delta}^2)}{(\nu^2-\bar{\Delta}^2)^2}\cos\left(4\pi\bar{\Delta}\tau_d\right)- 2
\end{equation}
\begin{equation}
    ff_{21}(\nu)=\frac{2\nu^2}{\nu^2-\bar{\Delta}^2}\cos\left(4\pi\bar{\Delta}\tau_d\right)- 2
\end{equation}
\begin{equation}
    ff_{22}(\nu)=\frac{2\nu\bar{\Delta}}{\nu^2-\bar{\Delta}^2}\sin\left(4\pi\bar{\Delta}\tau_d\right)
.\end{equation}

\begin{figure}[t]
    \centering
    \includegraphics[scale=0.6]{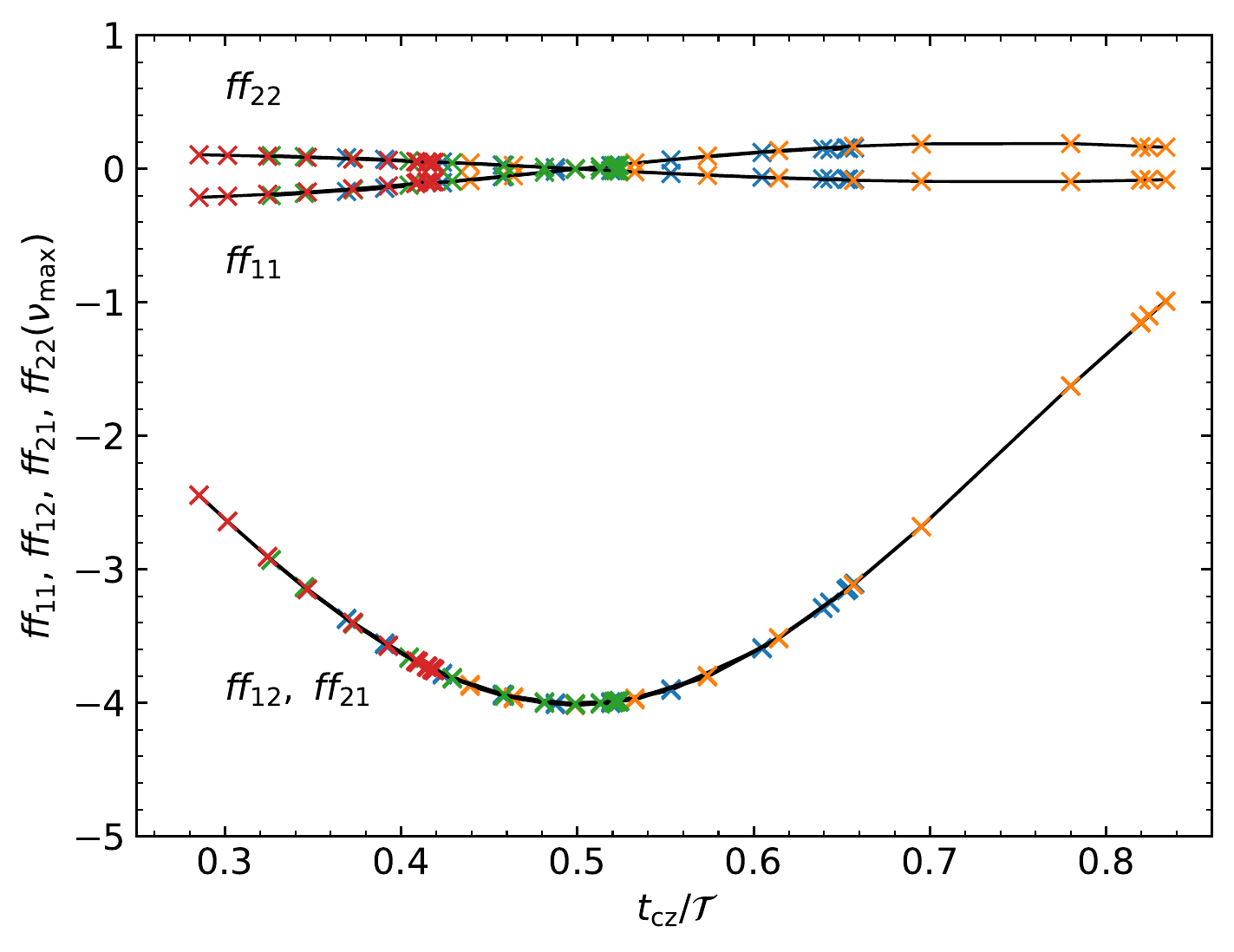}
    \caption{Value of the $ff$ functions at $\tilde{\nu}=\nu=\nu_\mathrm{max}$ for the same models as in Fig.~\ref{fig:7}, i.e.  with masses between $1.2$ and $1.5$~M$_\odot$ and $\xi_{PC}=2$.  }
   \label{fig:C1}
\end{figure}

Similarly to the ratios $r_{010}$ , the frequency-dependent terms $ff_{11}, ff_{22}$ are negligible compared to $ff_{12},ff_{21}$ , respectively. This is illustrated in Fig.~\ref{fig:C1}, which shows the $ff_{ij}$ coefficients. 

\medskip 

Again here as in Appendix A, we consider $\Delta /\nu<<1$ and $\Phi=2 \pi \Delta \tau_d$, then 

\begin{equation}
 ff_{12}(\nu) \sim ff_{21}(\nu)\approx   2 \cos\left(2 \Phi\right)- 2 =-4 \sin^2(\Phi)
.\end{equation}
\medskip

Moreover, for $2 \Phi \not=0$ (when $ff_{11}  = ff_{22}=0$), then

$$ff_{11}\approx -2ff_{22} \approx -4 \frac{\Delta}{\nu}$$

Furthermore, $\vert ff_{11}/ff_{12} \vert <<1$. 

Again taking $\Phi_t =\Phi \approx \pi/2$ , then 
$$ ff_{12}(\nu) \sim ff_{21}(\nu)\approx -4  $$
These types of approximated behaviour are shown in Fig.~\ref{fig:C1}.

\medskip
At $\nu\sim  \nu_{max}$ , we obtain 
\begin{equation}
    \begin{aligned}
          \Delta_2\nu_{n,l,cz}\approx &a_1(\tau) [ 
       ff_{12}(\nu)\sin\left(4\pi\nu\tau_d+2\phi\right)]\\
          &+a_2(\tau)  [ff_{21}(\nu)\cos\left(4\pi\nu\tau_d+2\phi\right) ]\\
         \approx &  -4  [ a_1(\tau) 
        \sin\left(4\pi\nu\tau_d+2\phi\right)]\\
          &+a_2(\tau) \cos\left(4\pi\nu\tau_d+2\phi\right) ]\\
    \end{aligned}
.\end{equation}

\end{appendix}
\end{document}